\documentclass[lettersize,journal]{IEEEtran}
\usepackage{bm} %
\usepackage{cite}
\usepackage{amsmath,amssymb,amsfonts}
\usepackage{algorithmic}
\usepackage{graphicx}
\usepackage{textcomp}
\usepackage{xcolor}

\usepackage{amsmath,graphicx}
\usepackage{amsmath,amsfonts}
\usepackage{algorithmic}
\usepackage{algorithm}
\usepackage{array}
\usepackage{hyperref}
\usepackage[caption=false,font=normalsize,labelfont=sf,textfont=sf]{subfig}
\usepackage{textcomp}
\usepackage{stfloats}
\usepackage{url}
\usepackage{verbatim}
\usepackage{graphicx}
\usepackage{cite}
\usepackage{bm}
\usepackage{multirow}
\usepackage{amssymb}
\usepackage{comment}
\usepackage{mathtools}
\usepackage{booktabs}
\usepackage{balance}
\usepackage{flushend}
\usepackage{siunitx}
\usepackage{xcolor}
\usepackage{arydshln}
\usepackage{enumitem}
\AtBeginDocument{
  \abovedisplayskip     =1\abovedisplayskip
  \abovedisplayshortskip=1\abovedisplayshortskip
  \belowdisplayskip     =1\belowdisplayskip
  \belowdisplayshortskip=1\belowdisplayshortskip
}
\usepackage{subcaption}

\def\R{{\mathbb R}}
\def\E{{\mathcal E}}
\def\D{{\mathcal D}}
\def\bandstart{g_{k}^{(s)}}
\def\bandend{g_{k}^{(e)}}

\begin{document}

\title{Input-Adaptive Spectral Feature Compression \\by Sequence Modeling for Source Separation}

\author{Kohei Saijo,~\IEEEmembership{Student Member,~IEEE,} Yoshiaki Bando,~\IEEEmembership{Member,~IEEE
\thanks{
Manuscript received xxx xxx, 2025; revised xxx xxx, 2025; accepted xxx xxx, 2025.
Date of publication xxx xxx, 2025; date of current version xxx xxx, 2025.
This work was supported in part by the JST FOREST No. JPMJFR232Y.
The associate editor coordinating the review of this manuscript and approving it for publication was Prof. xxxx xxxx.
\textit{(Corresponding author: Kohei Saijo)}   }
\thanks{
Kohei Saijo is with the Department of Communications and Computer Engineering, Waseda University, Tokyo 162-0042,
Japan (e-mail: saijo@pcl.cs.waseda.ac.jp)}%
}
\thanks{Kohei Saijo and Yoshiaki Bando are with the National Institute of Advanced Industrial Science and Technology (AIST), Tokyo, 135-0064, Japan (e-mail: y.bando@aist.go.jp).}
\thanks{Digital Object Identifier XX.XXXX/TASLPPRO.2025.XXXXXXX}}

\markboth{Journal of \LaTeX\ Class Files,~Vol.~14, No.~8, August~2021}%
{Shell \MakeLowercase{\textit{et al.}}: A Sample Article Using IEEEtran.cls for IEEE Journals}

\maketitle

\allowdisplaybreaks

\begin{abstract}

Time-frequency domain dual-path models have demonstrated strong performance and are widely used in source separation. 
Because their computational cost grows with the number of frequency bins, these models often use the band-split (BS) module in high-sampling-rate tasks such as music source separation (MSS) and cinematic audio source separation (CASS).
The BS encoder compresses frequency information by encoding features for each predefined subband.
It achieves effective compression by introducing an inductive bias that places greater emphasis on low-frequency parts.
Despite its success, the BS module has two inherent limitations: (i) it is not input-adaptive, preventing the use of input-dependent information, and (ii) the parameter count is large, since each subband requires a dedicated module.
To address these issues, we propose Spectral Feature Compression (SFC).
SFC compresses the input using a single sequence modeling module, making it both input-adaptive and parameter-efficient.
We investigate two variants of SFC, one based on cross-attention and the other on Mamba, and introduce inductive biases inspired by the BS module to make them suitable for frequency information compression.
Experiments on MSS and CASS tasks demonstrate that the SFC module consistently outperforms the BS module across different separator sizes and compression ratios.
We also provide an analysis showing that SFC adaptively captures frequency patterns from the input.

\end{abstract}
\section{Introduction}
\label{sec:intro}

\IEEEPARstart{A}{udio} source separation has achieved high performance through the advancement of neural networks.
Various networks have been proposed, including those based on time-frequency (TF) masking~\cite{dc,pit} and time-domain separation networks (TasNets)\cite{tasnet, convtasnet, dprnn, dptnet, sepformer, mossformer2}.
Compared with these approaches, which embed features at the time-frame level, recent studies have demonstrated that TF-domain dual-path models, which embed features at the TF-bin level and perform sequence modeling along the temporal and frequency dimensions alternately, achieve particularly strong performance~\cite{tfpsnet, tfgridnet, tflocoformer, tfcrossnet}.
These models achieve superior performance across a wide range of separation tasks, including speech enhancement and separation~\cite{tfgridnet,tflocoformer,tfcrossnet}, music source separation (MSS)~\cite{bsrnn,bsroformer}, and cinematic audio source separation (CASS)~\cite{bandit}.
However, since TF-domain dual-path models preserve the full frequency resolution of the input, they incur higher computational cost as the number of frequency bins increases.
This is particularly problematic in tasks such as MSS and CASS, where high sampling rates result in a large number of bins.

\begin{figure}[t]
\centering
\centerline{\includegraphics[width=1.0\linewidth]{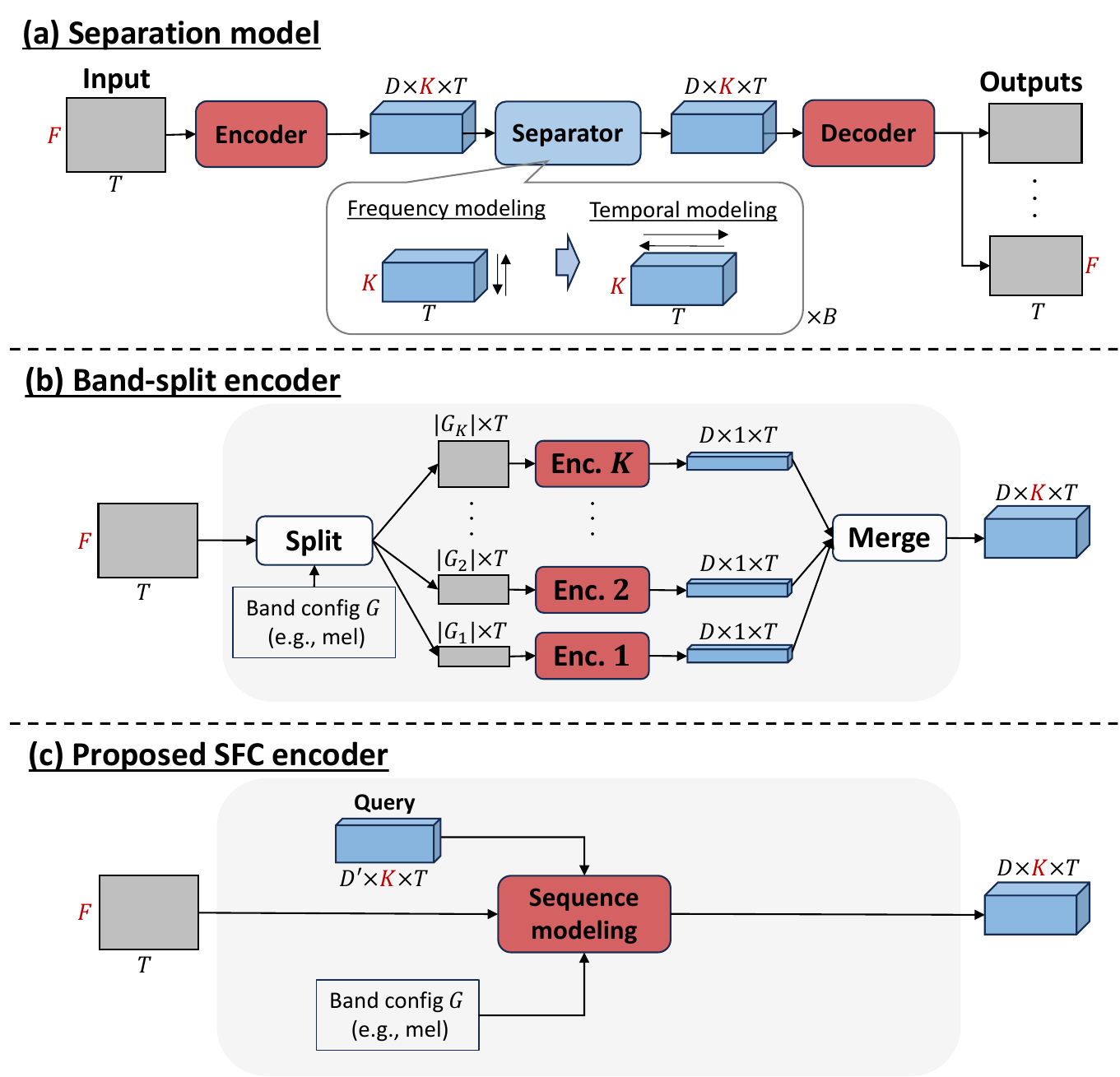}}
\caption{
Overview of (a) TF-domain dual-path separation models, (b) band-split (BS) encoder, and (c) proposed spectral feature compression (SFC) encoder.
The BS module divides an input spectrogram with $F$ frequency bins into $K$ subband spectrograms based on a predefined band configuration (e.g., mel) and processes them with $K$ different \textit{sub-encoders} to compress the inputs into $K$ subband features.
The proposed SFC also compresses the input into $K$ subband features, but does so with a single sequence modeling module using $K$ queries, making it both input-adaptive and parameter-efficient.
}
\label{fig:overview}
\end{figure}
To reduce computational cost, the band-split (BS) module~\cite{bsrnn} is often employed as the encoder and decoder in these tasks~\cite{simo_stereo_bsrnn, bsrnn_se, bsroformer, mamba2_mss, urgent2024, tuss, urgent2025_rank1st, mel_roformer, bandit}. 
The BS encoder splits the TF spectrogram into multiple subband spectrograms, encodes each subband with a dedicated \textit{sub-encoder}, and then merges the features (Fig.~\ref{fig:overview} (b)).
The decoder analogously consists of multiple \textit{sub-decoders}, each of which estimates the separated signals for its corresponding subband.
Motivated by the general assumption that lower frequencies have higher power and contain richer information, the BS module assigns narrower bandwidths to lower frequencies and wider bandwidths to higher frequencies.
Such band configurations are either handcrafted~\cite{bsrnn, simo_stereo_bsrnn, bsrnn_se, bsroformer, mamba2_mss, urgent2024, tuss, urgent2025_rank1st} or designed using psychoacoustical knowledge, such as mel scale~\cite{mel_roformer, bandit}.
Such an inductive bias enables the BS module to compress frequency information effectively.
Still, we consider that the BS module has room for improvement in the following two aspects:
(i) The encoding and decoding processes are not input-adaptive, preventing the module from exploiting input-dependent frequency patterns.
(ii) Each sub-encoder and sub-decoder requires its own parameters, leading to a large overall parameter count.

In contrast, other fields such as computer vision have explored input-adaptive feature compression using a single sequence-modeling module~\cite{perceiver_io, bimba}.
For example, Perceiver IO~\cite{perceiver_io} compresses input sequences of arbitrary length into sequences of fixed length by cross-attention~\cite{transformer} using learnable queries.
More recently, Mamba-based~\cite{mamba} sequence compression has been introduced for video processing, where queries are inserted at fixed intervals into the sequence, and the model is trained to output compressed representations when processing these queries~\cite{bimba}.
Compared with the BS module, these approaches (i) are input-adaptive, and (ii) require only a small number of parameters.
In addition, sequence modeling has the advantage of not restricting the receptive field, unlike the BS module where each sub-encoder and sub-decoder can only see a specific subband.
Although, to the best of our knowledge, such approaches have not yet been applied to compress frequency information in source separation, they can, in principle, be applied and overcome the limitations of the BS module.

Despite their potential, our preliminary experiments revealed that directly replacing the BS module with these methods results in substantially degraded performance.
We hypothesize that this is because these methods do not incorporate inductive biases and therefore fail to encode and decode frequency information as effectively as the BS module.

In this paper, we introduce input-adaptive feature compression methods designed for compressing frequency information.
The proposed approach, termed Spectral Feature Compression (SFC), builds on sequence modeling-based feature compression mentioned above~\cite{perceiver_io, bimba}, but incorporates psychoacoustically motivated inductive biases, inspired by the BS module, to capture frequency patterns effectively.
We investigate two variants of SFC: one with cross-attention and the other with recurrent modeling.
In the cross-attention variant, the inductive bias is incorporated via a carefully designed positional bias, whereas in the recurrent one, the inductive bias is introduced through the positions where queries are inserted.
The proposed SFC improves performance while demanding a much smaller number of parameters compared to the BS module.
We evaluate SFC on MSS and CASS tasks across different separator sizes and compression ratios.
In addition, we conduct extensive ablation studies to examine the effects of inductive bias and receptive field size.
Finally, we analyze the input adaptiveness of SFC by visualizing the attention weights.
The source code and pre-trained models are available at \url{https://github.com/b-sigpro/spectral-feature-compression}.

The rest of the paper is organized as follows.
Section~\ref{sec:background} describes TF-domain dual-path models and the BS module.
Section~\ref{sec:proposed_methods} introduces the proposed SFC with its motivation.
Section~\ref{sec:setup} explains the experimental setups, and Section~\ref{sec:results} reports the experimental evaluation and analysis.
Section~\ref{sec:conclusion} concludes this study with future directions.

\section{Background}
\label{sec:background}

In this section, we describe the background of this study, TF-domain dual-path models, typical encoder and decoder implementations, and the band-split module.

\subsection{TF-domain dual-path modeling}
\label{ssec:tf_domain_dual_path_modeling}

TF-domain dual-path models have been widely used in recent studies due to their strong performance~\cite{tfpsnet, tfgridnet,tflocoformer}.
These models generally follow an encoder–separator–decoder framework, and apply sequence modeling in both the frequency and temporal dimensions alternately (Fig.~\ref{fig:overview}, (a)).

Let the $M$-channel mixture\footnote{While $M$ can algorithmically be any number, we only consider monaural ($M=1$) or stereo ($M=2$) cases in this work.} in the STFT domain be denoted as $\bm{X} \in \R^{2M \times F \times T}$, where $F$ and $T$ are the numbers of frequency bins and time frames, respectively, and the factor of 2 corresponds to the real and imaginary (RI) components.
The mixture is encoded by an encoder $\E$, resulting in a TF bin-level feature, $\bm{Z} \in \R^{D \times F \times T}$, where $D$ is the feature dimension.

The core separator alternates between frequency and temporal modeling paths (see Fig.~\ref{fig:overview} (a)). 
In frequency modeling, the feature $\bm{Z}$ is viewed as a stack of $T$ arrays of shape $D \times F$, obtained by permuting $\bm{Z}$ into $T \times D \times F$. 
Each time frame, represented as a $D$-dimensional feature sequence of length $F$, is processed by a sequence modeling module, such as a recurrent~\cite{tfgridnet} or Transformer-based block~\cite{tflocoformer}.
In temporal modeling, $\bm{Z}$ is instead viewed as a stack of $F$ arrays of shape $D \times T$, and sequence modeling is applied along the temporal dimension.
These two processes are alternated multiple times, which enables interactions among all TF bins.

The separator output is passed to a decoder $\D$ to estimate separated signals $\hat{\bm{Y}} \in \R^{N \times 2M \times F \times T}$, where $N$ is the number of sources.
Several strategies exist to estimate separated signals, such as directly predicting the RI components, known as complex spectral mapping~\cite{complex_spectral_mapping}. In this work, following prior studies on MSS and CASS~\cite{bsrnn, bandit}, we configure the decoder to estimate complex masks~\cite{complex_mask}.
The separated signals are then obtained by multiplication of the estimated mask $\hat{\bm{M}} \in \R^{N \times 2M \times F \times T}$ with the mixture.

\subsection{Typical encoder and decoder design}
\label{ssec:conv2d_enc_dec}
In typical TF-domain dual-path models, the encoder $\E$ consists of normalization and a two-dimensional convolution layer, to produce a feature representation  $\bm{Z} \in \R^{D \times F \times T}$.
The decoder $\D$ analogously consists of a two-dimensional transpose convolution layer\footnote{Note that the operation does not necessarily have to be a transpose convolution, but can also be a standard convolution, since the stride is typically set to 1. However, as existing models often employ transpose convolution~\cite{tfgridnet,tflocoformer}, we adopt this configuration in our work.} to estimate $N$ separated signals.

Since TF-domain dual-path models apply sequence modeling to the feature $\bm{Z}$, which preserves the full frequency resolution of the input, their computational cost increases with the number of frequency bins $F$. 
Consequently, they become computationally expensive in tasks such as MSS or CASS, where data with high sampling rates (e.g., 44.1 kHz) are used, and the FFT size is set to a large value (e.g., 2048). 

\subsection{Band-split module}
\label{ssec:band_split_module}

To alleviate the computational burden of the separator, the band-split (BS) module is often employed in these tasks~\cite{bsrnn}.

The BS encoder maps the input to a $D$-dimensional feature, similarly to the encoder of TF-domain dual-path models, but simultaneously compresses the frequency dimension.
Specifically, while the encoder of TF-domain dual-path models is defined as a function $\E: \bm{X} \in \R^{2M \times F \times T} \mapsto \bm{Z} \in \R^{D \times F \times T}$ as described in Section~\ref{ssec:tf_domain_dual_path_modeling}, the BS encoder is defined as $\E: \bm{X} \in \R^{2M \times F \times T} \mapsto \bm{Z} \in \R^{D \times K \times T}$, where $K$ is the number of bands.
The decoder, in contrast, is designed to recover the original frequency resolution, $\D: \bm{Z} \in \R^{D \times K \times T} \mapsto \bm{Y} \in \R^{N \times 2M \times F \times T}$.
By setting $K < F$, the encoder compresses the frequency information, which reduces the computational cost of the separator.
In practice, $F$ is typically 1025 in MSS and CASS, and $K$ is set to much smaller values, such as 64.

An overview of the BS encoder is shown in Fig.~\ref{fig:overview} (b).
It splits the input spectrogram $\bm{X}$ into $K$ predefined subbands and encodes each using a dedicated \textit{sub-encoder}. 
Let $G_k = [\bandstart, \bandend]$ be the start and end index of the frequency bin at $k$-th band $(1 \leq \bandstart \leq \bandend \leq F, ~k=1, \dots, K)$ and the number of bins in the band $|G_k|:= \bandend - \bandstart$.
We also define a subband spectrogram $\bm{X}_{\bandstart: \bandend} := [\bm{X}_{\bandstart}, ..., \bm{X}_{\bandend}] \in \R^{2M \times |G_k| \times T}$.
Each subband is viewed as a matrix with a shape of ${2M|G_k| \times T}$ and encoded as $\bm{Z}_k = \E_{k}(\bm{X}_{\bandstart: \bandend}) \in \R^{D \times T}$, where $\E_{k}$ is the $k$-th sub-encoder.
The merged feature input to the separator $\bm{Z} \in \R^{D \times K \times F}$ is obtained by concatenating $K$ subband features.
Each sub-encoder typically consists of a normalization layer and a linear layer~\cite{bsrnn}.

The decoder analogously consists of $K$ \textit{sub-decoders} $\D_k$, which estimate the RI components of the mask for each subband from the corresponding feature: $\hat{\bm{M}}_{\bandstart: \bandend} = \D_{k}(\bm{Z}_{k})$, where $\hat{\bm{M}}_{\bandstart: \bandend} := [\hat{\bm{M}}_{\bandstart}, ..., \hat{\bm{M}}_{\bandend}] \in \R^{N \times 2M \times |G_k| \times T}$.
Each sub-decoder is typically implemented as a multi-layer perceptron (MLP), consisting of a normalization layer, two linear layers with $\tanh$ activations, and another linear layer with a gated linear unit (GLU) activation~\cite{bsrnn}.
By merging $\hat{\bm{M}}_{\bandstart: \bandend}$ for all $k$, we obtain $\hat{\bm{M}} \in \R^{N \times 2M \times F \times T}$.
Note that the bands can overlap.
In such a case, the final mask $\hat{\bm{M}}$ is obtained by taking the (weighted) sum of subband masks $\hat{\bm{M}}_{\bandstart:\bandend}$ across overlapping regions. 
For details of the weighted sum scheme, please refer to Section III-E of~\cite{bandit}.

To effectively compress the frequency information, the BS module introduces an inductive bias that places greater emphasis on low-frequency information, which is achieved by setting the bandwidth $|G_k|$ to be smaller at lower frequencies and larger at higher frequencies.
The design is based on the general assumption that lower frequencies have more power and contain richer information, thereby enabling effective compression of frequency information.
While such band configurations have originally been handcrafted~\cite{bsrnn}, recent studies have proposed psychoacoustically motivated designs such as the mel scale or the Bark scale~\cite{mel_roformer, bandit} and observed performance improvements.
Among them, the configuration based on the 12-tone equal temperament (12-TET) Western musical scale, hereafter referred to as the \texttt{Musical} scale, has been shown to yield superior performance.
In this work, we use this \texttt{Musical} scale to define the bands $G_k$.
For details of the \texttt{Musical} scale, please refer to~\cite{bandit}.
\section{Proposed Methods}
\label{sec:proposed_methods}

\begin{figure*}[h]
    \includegraphics[width=\textwidth]{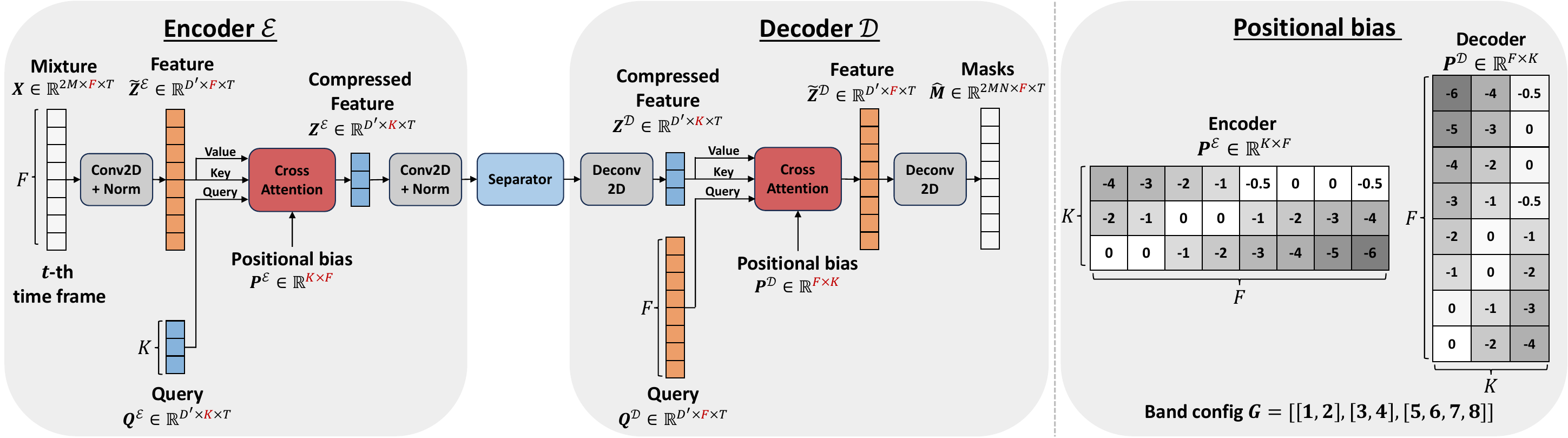}

     \caption{
        Detailed architecture of the proposed SFC using cross-attention (SFC-CA).
        The SFC-CA encoder compresses input spectral features of length $F$ into compressed features of length $K$ through cross-attention with randomly initialized learnable queries of length $K$. The decoder uncompresses the features in a similar manner, but with queries of length $F$. To introduce a psychoacoustically motivated inductive bias, similarly to the BS module, we design a positional bias (right; Sec.~\ref{sssec:crossattn_with_inductive_bias}) based on the band definition $G$, which assigns higher values to frequency bins $f$ included in the $k$-th band.
        The vertical boxes represent the data at a specific time frame $t$, whereas the shapes of the variables refer to those of the entire tensor across all $T$ time frames.
     }
     \label{fig:sfc_ca}
\end{figure*}

Despite its effectiveness, the BS module has room for improvement in the following aspects.
(i) It is not input-adaptive. While each input spectrogram has different frequency patterns, which may enable more effective compression, the BS module cannot leverage such input-dependent information.
(ii) Since distinct parameters are used to process each subband, the total parameter count becomes large. For example, under the practical configuration used in this study (the \texttt{Musical} split with $K=64$ bands), the combined parameter count of the encoder and decoder is more than twice that of the separator.

Given these considerations, we propose Spectral Feature Compression (SFC), an input-adaptive and parameter-efficient feature compression method based on sequence modeling.
Instead of splitting the input spectrogram and using $K$ sub-encoders, we perform encoding with a single sequence modeling module using $K$ queries.
By designing the $k$-th query to embed information primarily from the $k$-th band $G_k$, we introduce an inductive bias similar to that of the BS module, making SFC suitable for compressing frequency information.
We explore two variants, one with cross-attention (SFC-CA) and the other with recurrent models (SFC-Mamba).
Although (bidirectional) attention often outperforms recurrent modeling in many fields, recurrent models have also achieved strong performance in source separation~\cite{tfgridnet}, where local modeling plays an important role.
This suggests that both approaches are promising, which motivates us to explore both variants.

Section~\ref{ssec:overview} provides an overview of the SFC module.
Sections~\ref{ssec:cross_attention} and~\ref{ssec:reccurent_modeling} describe its design, including the rationale behind the proposed architecture.
Section~\ref{ssec:relation_to_other_methods} reviews related methods in order to position the proposed SFC approach.

\subsection{Overview of proposed SFC}
\label{ssec:overview}

\begin{figure*}[h]
    \includegraphics[width=\textwidth]{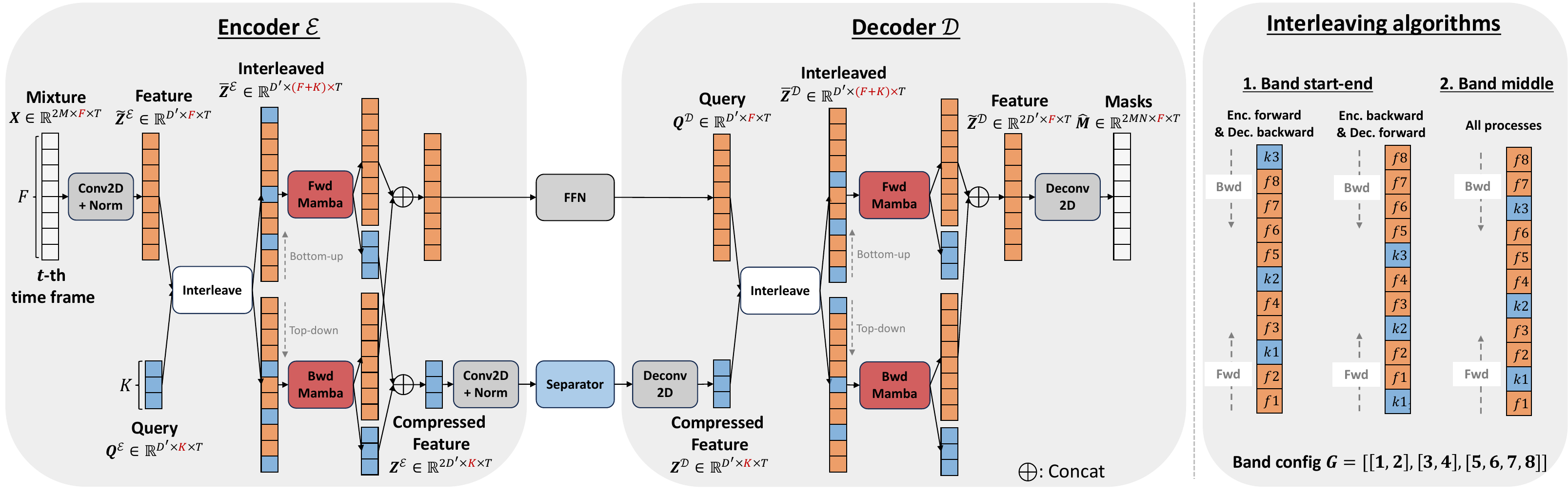}

     \caption{Detailed architecture of proposed SFC using recurrent models (SFC-Mamba). SFC-Mamba encoder first interleaves the input feature and query.
     Mamba is then applied, and outputs at the queries' position (blue boxes) are used as the compressed feature input to the separator.
     We apply Mamba bidirectionally; one scans the sequence in the normal order (forward Mamba) and the other does so in the reversed order (backward Mamba).
     The decoder also applies bidirectional Mamba to the interleaved sequence, but queries in the decoder are derived from the outputs at the features' position (orange boxes) of the encoder's Mamba.
     Interleaving algorithms (right) are designed to introduce a psychoacoustically motivated inductive bias, motivated by the band-split module (Section~\ref{sssec:interleaving_strategy}).
     The vertical boxes represent the data at a specific time frame $t$, whereas the shapes of the variables refer to those of the entire tensor across all $T$ time frames.
     }
     \label{fig:sfc_mamba}
\end{figure*}

As shown in Fig.~\ref{fig:sfc_ca} and Fig.~\ref{fig:sfc_mamba}, the encoder $\E$ of SFC mainly consists of the following three steps:
\begin{align}
  \label{eq:encoder}
    \tilde{\bm{Z}}^{\E} &= \mathrm{Norm}(\mathrm{Conv2D}(\bm{X})) \in \R^{D' \times F \times T}, \\    
    \bm{Z}^{\E} &= \mathrm{Seq}(\tilde{\bm{Z}}^{\E}, \bm{Q}^{\E}) \in \R^{D' \times K \times T}, \\
    \bm{Z} &= \mathrm{Norm}(\mathrm{Conv2D}(\bm{Z}^{\E})) \in \R^{D \times K \times T},
\end{align}
where $\mathrm{Norm}$ is root-mean-square (RMS) normalization~\cite{rmsnorm}, and $\mathrm{Conv2D}$ is two-dimensional convolution.
$\mathrm{Seq}$ is the input-adaptive sequence modeling-based feature compression, which
incorporates an inductive bias inspired by the BS module to capture frequency information well, as described in Sections~\ref{sssec:crossattn_with_inductive_bias} and~\ref{sssec:interleaving_strategy}.
$\bm{Q}^{\E} \in \R^{D' \times K}$ denotes the query used to compress the feature.
In summary, the SFC encoder first encodes the input TF spectrogram as in the TF-domain dual-path models, compresses the frequency dimension, and then applies $\mathrm{Conv2D}$ and $\mathrm{Norm}$ again.
Note that we compress the feature using $D' (<D)$ dimensional feature and then map it to $D$-dimensional feature to save the computational cost of $\mathrm{Seq}$, which is why we have two $\mathrm{Conv2D}$ and $\mathrm{Norm}$ operations.

The separator, which alternates sequence modeling along frequency and temporal dimensions as described in Section~\ref{ssec:tf_domain_dual_path_modeling}, receives the compressed feature $\bm{Z}$ and outputs a processed feature:
\begin{align}
  \label{eq:separator}
    \bm{Z} \xleftarrow{} \mathrm{Separator}(\bm{Z}) \in \R^{D \times K \times T}.
\end{align}

The decoder $\D$ is designed to recover the original frequency resolution and to be symmetric to the encoder $\E$:
\begin{align}
  \label{eq:decoder}
    \bm{Z}^{\D} &= \mathrm{Deconv2D}(\bm{Z}) \in \R^{D' \times K \times T}, \\    
    \tilde{\bm{Z}}^{\D} &= \mathrm{Seq}(\bm{Z}^{\D}, \bm{Q}^{\D}) \in \R^{D' \times F \times T}, \\
    \hat{\bm{M}} &= \mathrm{Deconv2D}(\tilde{\bm{Z}}^{\D}) \in \R^{N \times 2M \times F \times T},
\end{align}
where $\mathrm{Deconv2D}$ is the two-dimensional transpose convolution.
Hereafter, we omit the time dimension $T$ for simplicity, because $\mathrm{Seq}$ is applied to each frame independently.

\subsection{SFC with cross-attention (SFC-CA)}
\label{ssec:cross_attention}

\subsubsection{Basic architecture}
The overview of SFC-CA is shown in Fig.~\ref{fig:sfc_ca}.
The cross-attention block in the encoder $\E$ generally follows the original formulation in~\cite{transformer}, but incorporates several modifications to adapt it for spectral feature compression:
\begin{align}
  \label{eq:cross_attention_block}
    \bm{Q} &= \mathrm{Linear}(\bm{Q}^{\E}) \in \R^{D' \times K}, \\
    \bm{K} &= \mathrm{Linear}(\tilde{\bm{Z}}^{\E}) \in \R^{D' \times F},\\ 
    \bm{V} &= \mathrm{Linear}(\tilde{\bm{Z}}^{\E}) \in \R^{D' \times F}, \\ 
    \bm{Z}^{\E} &= \mathrm{CrossAttn}(\bm{Q}, \bm{K}, \bm{V}) \in \R^{D' \times K},\\ 
    \bm{Z}^{\E} &= \mathrm{FFN}(\bm{Z}^{\E})  + \bm{Z}^{\E} \in \R^{D' \times K},
\end{align}
where $\mathrm{Linear}$ is a single fully connected layer, and $\mathrm{FFN}$ is a two-layer feed-forward network (FFN) with internal feature dimension of $2D'$ using swish gated linear unit (SwiGLU) activation~\cite{swiglu}, which is often employed in recent Transformer models~\cite{llama2, tflocoformer}.
$\mathrm{CrossAttn}$ is multi-head cross-attention~\cite{transformer}, and is described below in detail.
Note that we do not use the skip connection after $\mathrm{CrossAttn}$.
The decoder follows the architecture of the encoder, but the length of the query $\bm{Q}^{\D}$ is set to $F$ to obtain the uncompressed feature with $F$ frequency bins.

\subsubsection{Cross-attention with inductive bias}
\label{sssec:crossattn_with_inductive_bias}

Unlike the normal cross-attention, the cross-attention in SFC is defined as:
\begin{align}
  \label{eq:cross_attention}
    \mathrm{CrossAttn}(\bm{Q}, \bm{K}, \bm{V}) := \mathrm{softmax}\left(\frac{\bm{Q}\bm{K}^\top}{\sqrt{D'}} + \gamma\bm{P}\right)\bm{V},
\end{align}
where $\bm{P}$ is the positional bias to introduce the inductive bias, and $\gamma$ is a slope value.
In standard cross-attention without positional bias $\bm{P}$, all key–query pairs are treated with equal importance.
In our case, this implies that all $K$ queries assign the same importance to all frequency bins, thereby imposing no inductive bias.
In contrast, we introduce an inductive bias, analogous to the BS module, by enforcing the $k$-th query to prioritize the frequency bins in the $k$-th band $G_k$.
Specifically, using the pre-defined band configuration $G_k$, we define the positional bias for the encoder $\bm{P}^{\E} \in \R^{K \times F}$ as:
\begin{align}
    \label{eq:position_bias}
    P^{\E}_{k,f} &= 
    \begin{cases}
    \frac{|(\bandend + \bandstart) / 2 - f|}{|G_k|}, & \text{if $\bandstart \leq f \leq \bandend$,} \\
    \bandend - f, & \text{if $\bandend < f$,} \\
    f - \bandstart, & \text{if $\bandstart > f$,}
    \end{cases}
\end{align}
and that for the decoder $\bm{P}^{\D} \in \R^{F \times K}$ as:
\begin{align}
    \label{eq:position_bias_dec}
    P^{\D}_{f,k} &= P^{\E}_{k,f}.
\end{align}
Eq.(\ref{eq:position_bias}) implies that (i) if the $f$-th frequency bin lies within the $k$-th band $G_k$, the positional bias is set to a small value between -1 and 0\footnote{While setting $P_{k,f}=0$ for all $f$ in the $k$-th band (i.e., treating them with equal importance) may be more intuitive, the definition in Eq.(\ref{eq:position_bias}) gave slightly better performance in preliminary experiments with fixed $\bm{P}$.}, and (ii) otherwise, the positional bias decreases in proportion to the distance from the edge of the band.
In other words, for a given band $G_k$, the farther the $f$-th frequency bin is from $G_k$, the lower (more negative) value is assigned, leading it to be progressively ignored in cross-attention.
Fig~\ref{fig:pos_bias} (top) visualizes the encoder's positional bias $\bm{P}^{\E}$.
To make the decoder symmetric to the encoder, we use the transposed matrix $(\bm{P}^{\E})^{\top}$ as $\bm{P}^{\D}$.

While Eq.~(\ref{eq:position_bias}) was found to be a reasonable formulation for incorporating an inductive bias in cross-attention through our preliminary experiments, we acknowledge three aspects that leave room for improvement:
(i) the positional bias is symmetric with respect to the band center,
(ii) the positional bias values decay linearly both inside and outside the band, and
(iii) the decay slope is fixed to $\gamma$.
To address these limitations, we explore the following two strategies:
\begin{itemize}
    \item \textbf{Learnable $\gamma$}: The slope value $\gamma$ is initialized to 1 and treated as a learnable parameter. This strategy addresses the third limitation.

    \item \textbf{Learnable $\bm{P}$}: The positional biases $\bm{P}^{\E}$ and $\bm{P}^{\D}$ are initialized using Eqs.~(\ref{eq:position_bias}) and (\ref{eq:position_bias_dec}), respectively, and then made learnable.
    Since all elements of the positional bias can be directly optimized, this strategy has the potential to alleviate all three limitations.  
\end{itemize}
Note that each head in cross-attention has its own $\gamma$ and $\bm{P}$.

\subsubsection{Query}
\label{sssec:learnable_query}
SFC-CA uses learnable queries, which are randomly initialized and jointly optimized with the separator, in both the encoder and the decoder.
The shape of the queries in the encoder, $\bm{Q}^{\E}$, and decoder, $\bm{Q}^{\D}$, are defined as $D' \times K$ and $D' \times F$, respectively, without including the time-frame dimension.
They are repeated $T$ times in each forward pass.
Hereinafter, we refer to this type of queries as \texttt{Learnable} (in contrast to the alternative defined in Section~\ref{sssec:input_query}).

\subsection{SFC with recurrent modeling (SFC-Mamba)}
\label{ssec:reccurent_modeling}

\subsubsection{Basic architecture}
The overview of SFC-Mamba is shown in Fig.~\ref{fig:sfc_mamba}.
The encoder uses Mamba~\cite{mamba} to compress the frequency information, as formulated below:
\begin{align}
  \label{eq:mamba_block}
    \bar{\bm{Z}}^{\E}_{\mathrm{fwd}}, \bar{\bm{Z}}^{\E}_{\mathrm{bwd}} &\xleftarrow{} \mathrm{Interleave}(\tilde{\bm{Z}}^{\E}, \bm{Q}^{\E}), \\ 
    \bm{Z}^{\E}_{\mathrm{fwd}}, \bm{Q}^{\E}_{\mathrm{fwd}} &\xleftarrow{} \mathrm{Mamba}_{\mathrm{fwd}}(\bar{\bm{Z}}^{\E}_{\mathrm{fwd}}), \\ 
    \bm{Z}^{\E}_{\mathrm{bwd}}, \bm{Q}^{\E}_{\mathrm{bwd}} &\xleftarrow{} \mathrm{Mamba}_{\mathrm{bwd}}(\bar{\bm{Z}}^{\E}_{\mathrm{bwd}}), \\ 
    \bm{Z}^{\E} &= \mathrm{Concat}(\bm{Z}^{\E}_{\mathrm{fwd}}, \bm{Z}^{\E}_{\mathrm{bwd}}). 
\end{align}
First, the feature $\tilde{\bm{Z}}^{\E}$ and the query $\bm{Q}^{\E}$ are interleaved to form a single sequence $\bar{\bm{Z}}^{\E} \in \R^{D' \times (F + K)}$ (detailed in Section~\ref{sssec:interleaving_strategy}).
The sequence is then processed by $\mathrm{Mamba}$.
As illustrated in Fig.\ref{fig:sfc_mamba}, we apply Mamba bidirectionally~\cite{vision_mamba, mamba_in_speech}; the forward Mamba ($\mathrm{Mamba}_{\mathrm{fwd}}$) scans the sequence from low to high frequencies, and the backward Mamba ($\mathrm{Mamba}_{\mathrm{bwd}}$) does so in a reversed order from high to low frequencies.
After applying bidirectional Mamba, the outputs at the \textit{query positions} are taken as the compressed features $\bm{Z}^{\E}_\mathrm{fwd}$ and $\bm{Z}^{\E}_\mathrm{bwd} \in \R^{D' \times K}$.
The outputs at the \textit{feature positions}, denoted as $\bm{Q}_\mathrm{fwd}^{\E}$ and $\bm{Q}_\mathrm{bwd}^{\E} \in \R^{D' \times F}$, are later used as queries in the decoder (detailed in Section~\ref{sssec:input_query}).
Finally, the outputs of the forward and backward passes are concatenated at the feature dimension to obtain $\bm{Z}^{\E} \in \R^{2D' \times K}$.
The decoder is designed to be symmetric to the encoder so that we can obtain an uncompressed feature with length $F$, similarly to the decoder of SFC-CA.

\subsubsection{Interleaving strategy to incorporate inductive bias}
\label{sssec:interleaving_strategy}

Similarly to SFC-CA, we introduce an inductive bias to make SFC-Mamba suitable for frequency information compression.
Because recurrent models are inherently dependent on sequence order, we incorporate the inductive bias by carefully configuring the query insertion positions. Specifically, the query positions are determined based on the band definitions $G_k$.
We propose two strategies, \texttt{Band start-end} and \texttt{Band middle}, as illustrated in Fig.~\ref{fig:sfc_mamba} (right).

In the \texttt{Band start-end} strategy, the $k$-th query is inserted at the end of the $k$-th band, so that Mamba can first scan the features within each band and then output the compressed feature upon reaching the query.
Let $I(k)$ denote the index where the $k$-th query $\bm{Q}^{\E}_{k}$ is inserted.
In the encoder, $I(k)$ in the forward Mamba is defined as:
\begin{align}
  \label{eq:bs_end_forward}
     I(k) := \bandend + k,
\end{align}
where $k$ is added because $k-1$ queries have already been inserted when the $k$-th query is placed.
In contrast, for the backward Mamba, the $k$-th query is inserted at the start of the $k$-th band, since the sequence is scanned in the reverse order:
\begin{align}
  \label{eq:bs_end_backward}
     I(k) := \bandstart + k.
\end{align}
In the decoder, however, the model must first scan the $k$-th compressed feature $\bm{Z}^{\D}_{k}$ to decode frequency information contained in the $k$-th band.
Accordingly, in the forward Mamba, we insert $\bm{Z}^{\D}_{k}$ at the start of the $k$-th band, which is the same strategy as that of the encoder’s \textit{backward} Mamba (Eq.~(\ref{eq:bs_end_backward})).
Conversely, in the backward Mamba, we insert $\bm{Z}^{\D}_{k}$ at the end of the $k$-th band, which corresponds to Eq.~(\ref{eq:bs_end_forward}).

In the \texttt{Band middle} strategy, the $k$-th query is inserted at the middle of the $k$-th band $G_k$:
\begin{align}
\label{eq:bs_middle}
I(k) := \Big\lfloor \tfrac{\bandstart + \bandend}{2} \Big\rfloor + k.
\end{align}
Note that we can use Eq.~(\ref{eq:bs_middle}) in both the forward and backward Mambas, and both in the encoder and decoder.

Whereas the \texttt{Band start-end} strategy enforces both forward and backward Mambas to compress all the information in each band, the \texttt{Band middle} strategy allows the two Mambas to compress only half of the band information.
While the former may better preserve the input information, the latter may be easier to learn.

\subsubsection{Query}
\label{sssec:input_query}
In SFC-Mamba, we adaptively compute the queries from the input feature (referred to as \texttt{Adaptive} queries).
Specifically, the queries in the encoder $\bm{Q}^{\E}$ are obtained by calculating the weighted average of the encoder's internal feature $\tilde{\bm{Z}}^{\E}$ in each band $G_k$:
\begin{align}
  \label{eq:average_query_enc}
    \bm{Q}^{\E}_{k} = \sum_{f=\bandstart}^{\bandend} w_{f}\tilde{\bm{Z}}^{\E}_{f},
\end{align}
where the weight $w_f$ is a learnable scalar.
The decoder's query $\bm{Q}^{\D}$, in contrast, is obtained by applying an $\mathrm{FFN}$ block to the outputs of the encoder's Mamba (Eq.~(\ref{eq:mamba_block})):
\begin{align}
  \label{eq:average_query_dec}
    \bm{Q}^{\D}_{f} = \mathrm{FFN}(\mathrm{Concat}(\bm{Q}^{\E}_{\mathrm{fwd}}, \bm{Q}^{\E}_{\mathrm{bwd}})).
\end{align}

The key difference between \texttt{Learnable} and \texttt{Adaptive} queries lies in whether the queries contain input information.
In cross-attention, the query is used solely for computing the attention matrix, and it is thus expected to work effectively even without input information.
In contrast, in recurrent models, the query is directly scanned to obtain the compressed feature or the uncompressed feature, and thus may benefit from queries that include input information.
For this reason, we use the \texttt{Adaptive} query by default in SFC-Mamba.

\subsection{Relation to other methods}
\label{ssec:relation_to_other_methods}

This section reviews several related works to position our methods.

\subsubsection{Perceiver IO}
\label{sssec:perceiver_io}
SFC-CA is built upon the framework of Perceiver IO~\cite{perceiver_io}.
Perceiver IO compresses sequences of arbitrary length into fixed-length representations using cross-attention, but without incorporating positional bias. 
One of our contributions is to redesign such cross-attention-based encoders and decoders so that they are suitable for frequency information compression. 
Through our experiments, we demonstrate that introducing the proposed positional bias is crucial for effective spectral feature compression.

\subsubsection{ALiBi}
\label{sssec:alibi}
The design of our positional bias is inspired by Attention with Linear Biases (ALiBi)~\cite{alibi}.
In ALiBi, the difference between the query and key indices is used as a positional bias, such that in self-attention, keys closer to the query receive higher attention weights.
Motivated by this idea, we introduce an inductive bias that captures the input frequency pattern by defining the positional bias as a function of the distance between each frequency bin and its corresponding band $k$.
Furthermore, we leverage the fact that the shape of the positional bias is always fixed in our case and configure the positional bias to be learnable, thereby avoiding the issue of symmetry in the positional bias.\footnote{ALiBi was originally designed for causal decoder-only Transformers, where symmetry in the positional bias does not pose a problem. However, symmetric positional bias may not be optimal in non-causal models like SFC.}

\subsubsection{BIMBA}
\label{sssec:bimba}
SFC-Mamba is based on Bidirectional Interleaved Mamba for Better Answers (BIMBA)~\cite{bimba}, which we extend to SFC-Mamba.
BIMBA compresses temporal sequences by inserting queries at fixed intervals and picking up the outputs at the query positions.
In contrast, SFC-Mamba introduces an inductive bias by carefully designing the query insertion strategies, making it more suitable for compressing frequency information.
Moreover, while BIMBA was originally designed for the video question answering task, which does not require decoding, we introduce a decoder that recovers uncompressed features from compressed representations, which allows us to adapt the model for source separation.
\section{Experimental setup}
\label{sec:setup}

This section describes the experimental setup, including the datasets, the configurations of the separator and encoder/decoder, and the details of training and evaluation.

\subsection{Datasets}
\label{ssec:datasets}

\subsubsection{MSS task}
MUSDB18-HQ~\cite{MUSDB18HQ} was used to evaluate performance on the MSS task. 
Each song is provided as a mixture of four stems (vocals, bass, drums, and other), and the goal is to separate the mixture into these four sources.
MUSDB18-HQ contains $100$ training songs and $50$ test songs, all sampled at $44.1$ kHz.
Following a conventionally adopted split, $86$ songs were used for training and $14$ for validation.
To remove silent regions, we applied the unsupervised source activity detection (SAD) method introduced in~\cite{bsrnn} to the training data, using a segment duration of $8$ seconds.

\subsubsection{CASS task}
Divide-and-remaster (DnR) dataset~\cite{dnr} was used for evaluating the models on the CASS task, where the goal is to separate a mixture into speech, music, and sound effect (SFX) stems.
The dataset contains 3406, 487, and 973 tracks in training, validation, and test sets, respectively.
All tracks are 1 minute long and sampled at 44.1 kHz.
We applied the same SAD method as in MUSDB18-HQ.

\subsection{Separator}
\label{ssec:separator}

In the separator, we use TF-Locoformer~\cite{tflocoformer} blocks, one of the state-of-the-art models in source separation\footnote{In preliminary experiments, we observed that TF-Locoformer with the BS module (BS-Locoformer) achieves performance comparable to that of the state-of-the-art model BS-Roformer~\cite{bsroformer} when certain training configurations are matched (see \texttt{B4} in Table~\ref{table:results_musdb}). We therefore adopted TF-Locoformer blocks for the separator, as they resulted in approximately twice the training speed compared with BS-Roformer in our experimental setting.}.
Each TF-Locoformer block consists of frequency modeling and temporal modeling sub-blocks, both based on multi-head self-attention and convolutional FFNs.
Please refer to~\cite{tflocoformer} for further architectural details. 
We basically apply the original TF-Locoformer block without any changes, but following~\cite{tflocoformer_nope}, we removed positional encoding.

We investigate two sizes of separators, small and medium-sized ones.
In the small configuration, the number of TF-Locoformer blocks is set to $B=4$, the embedding dimension to $D=96$, the hidden dimension in the FFNs to $C=128$, the kernel size in convolution to $K=8$, the stride of convolution to $S=1$, the number of attention heads to $H=4$, and the number of groups in group normalization to $G=4$.
In the medium one, we set $B=6$, $D=128$, $C=192$, $H=8$, and $G=8$, while keeping all other configurations unchanged.
Note that the notation here is consistent with that used in the original TF-Locoformer paper and is independent of the notation defined in other sections of this paper.
The small and medium separators contain around $5.0$ M and $15.0$ M parameters, respectively.

\subsection{Compared methods: encoder and decoder}
\label{ssec:enc_dec}

In the encoder and decoder, the default configuration below is used unless otherwise noted.
The STFT window size was set to 2048, and the hop size was set to 512.
For the band configuration $G$, all models employed the \texttt{Musical} split~\cite{bandit} (Section~\ref{ssec:band_split_module}) with $K=64$ bands.
In the SFC encoder and decoder, the kernel size and the stride of $\mathrm{Conv2D}$ and $\mathrm{Deconv2D}$ operations were set to 3 and 1, respectively.
We compare the following three encoders/decoders.

\textbf{BS module}: The BS module described in Section~\ref{ssec:band_split_module}.
Following previous studies~\cite{bandit}, the encoder consists of a normalization layer and a linear layer, while the decoder consists of a normalization layer and an MLP, where the inner feature dimension of the MLP was set to $4D$.
The encoder/decoder of the small model contained around 0.8 M / 28.8 M parameters, and that of the medium model had around 1.1 M / 29.6 M parameters, respectively.

\textbf{SFC-CA}: The cross-attention variant of the proposed SFC, described in Section~\ref{ssec:cross_attention}. 
\texttt{Learnable} query was used in both the encoder and decoder.
Based on preliminary experiments, by default, we fixed the slope $\gamma$ to 1, and configured the position bias $\bm{P}$ initialized with Eqs.~(\ref{eq:position_bias}) and (\ref{eq:position_bias_dec}) to be learnable.
The inner feature dimension of the module $D'$ was set to 64 and 96 for the small and medium models, respectively.
In both sizes, the number of attention heads $H$ was set to 4.
The encoder and decoder of the small model contained approximately 0.37 M and 0.43 M parameters, respectively, while those of the medium model contained approximately 0.48 M and 0.58 M parameters. 
Most of the parameters come from the positional biases $\bm{P}^{\E}$ and $\bm{P}^{\D}$, each of which has $F \times K \times H = 1025 \times 64 \times 4 \approx 0.26$ M parameters.

\textbf{SFC-Mamba}: The recurrent variant of the proposed SFC using Mamba, described in Section~\ref{ssec:reccurent_modeling}.
The state size of Mamba was set to 8, as no further performance improvement was observed when increasing it beyond this value.
Unlike the SFC-SA, \texttt{Adaptive} query was used.
By default, we use the \texttt{BS-middle} strategy when interleaving the sequences.
$D'$ is set to 32 and 48 for the small and medium models, respectively.
Note that, considering that the recurrent variant has two passes, forward and backward Mambas, each processing $D'$-dimensional features, we set $D'$ to half of that used in SFC-CA.
The encoder/decoder of the small model contained around 0.07 M / 0.06 M parameters, and that of the medium model had around 0.13 M / 0.11 M parameters, respectively.

\subsection{Training and evaluation details}
\label{ssec:train_eval_details}

\subsubsection{MSS task}
The models were trained for 900 epochs, each consisting of 110 training steps.
We used the AdamW optimizer~\cite{adamw} with a weight decay factor of 1e-2.
The learning rate was linearly increased from 0 to 1e-3 over the first 5000 training steps, kept constant until 550 epochs, and then decayed by a factor of 0.98 every two epochs. 
The input length was 6 seconds, and the batch size was set to 32. 
Gradient clipping was applied with a maximum gradient $L_2$-norm of 5. 
We used automatic mixed precision with bfloat16 precision.
During training, dynamic mixing (DM)~\cite{twostep, wavesplit} was performed at each step: a segment from each stem was randomly selected, RMS-normalized, scaled by a gain uniformly sampled from [-10, 10] dB, and then mixed.
Following~\cite{bsrnn}, each segment was dropped with a probability of 0.1 before mixing to simulate mixtures in which the target source is inactive.
The loss function was based on the negative thresholded SNR~\cite{mixit}, but was modified to accept zero signals as references, following the approach in~\cite{fuss}:
\begin{align}
    \label{eq:snr_loss}
    \mathcal{L}(y, \hat{y})  &= 
    \begin{cases}
     -10\log_{10}{\frac{||y||^2}{||y-\hat{y}||^2 + \tau||y||^2},} &\text{$||y||^2>0$,} \\
     -\alpha10\log_{10}{\frac{1}{||\hat{y}||^2 + \tau||x||^2},} &\text{$||y||^2=0$,}
    \end{cases}
\end{align}
where $x$, $y$, and $\hat{y}$ denote the mixture, the reference signal, and the estimated signal in the time domain, respectively.
The second term encourages the model to output zero signals when the reference is silent, and $\alpha$ is a weighting factor set to 0.1.

During inference, each song was segmented into 12-second chunks with a 6-second overlap, separated individually, and then reconstructed using overlap-add to obtain song-level results.
Note that we used longer segments than in training, as this has been shown to yield better performance~\cite{simo_stereo_bsrnn, tflocoformer_nope}.
For evaluation, we adopted track-level SNR (uSDR)\cite{usdr} and chunk-wise SDR (cSDR)\cite{csdr} as the metrics.
In the ablation study, we also evaluate the difference between the reference and estimated amplitude spectrograms, which we refer to as SpecSNR, to see the spectrogram-level reconstruction quality:
\begin{align}
    \label{eq:specsnr}
    \mathrm{SpecSNR}(Y, \hat{Y}) = 10\log_{10}\frac{|Y|^2}{(|Y| - |\hat{Y}|)^2},
\end{align}
where a higher value indicates better performance.

\subsubsection{CASS task}
We basically followed the training and evaluation setups used in MSS, but made several modifications as described below.
The models were trained for 150 epochs, with each epoch consisting of 1354 training steps.
The learning rate was warmed up using the same strategy as in MSS, kept constant until 75 epochs, and then decayed by a factor of 0.96 per epoch.
The source-dropping probability in DM was reduced to 0.05, considering that the number of stems in CASS is smaller than that in MSS.
For evaluation, we used SNR and scale-invariant signal-to-distortion ratio (SISDR)~\cite{le2019sdr}, both computed at the track level.
All other configurations were consistent with those in MSS.
\section{Results}
\label{sec:results}

\begin{table*}[t]
\centering
\sisetup{
detect-weight, %
mode=text, %
tight-spacing=true,
round-mode=places,
round-precision=2,
table-format=2.2,
table-number-alignment=center
}
\caption{
    cSDR and uSDR [dB] on MUSDB18-HQ test set.
    Numbers in `Encoder/Decoder` cells are number of bands.
    Prior work uses their ownband configurations, while models we trained (\texttt{A*} and \texttt{B*}) use \texttt{Musical} split.
    Results marked in grey are not directly comparable because ($\dagger$) validation split is not explicitly mentioned, and ($\ddagger$) uncommon validation split is used.
    \texttt{B4} is trained using the entire 100 songs for training while changing the input duration to 8 seconds.
}

\label{table:results_musdb}
\resizebox{\linewidth}{!}{
\begin{tabular}{lllc*{12}{S}}

\toprule

\multirow{2}{*}[-1.3ex]{\texttt{ID}} &\multirow{2}{*}[-1.3ex]{\shortstack{Model}} &\multirow{2}{*}[-1.3ex]{\shortstack{Encoder/Decoder}} &\multirow{2}{*}[-1.3ex]{\shortstack{Params.}} &\multicolumn{2}{c}{Vocals} &\multicolumn{2}{c}{Bass} &\multicolumn{2}{c}{Drums} &\multicolumn{2}{c}{Other} &\multicolumn{2}{c}{Average} \\
\cmidrule(lr){5-6}\cmidrule(lr){7-8}\cmidrule(lr){9-10}\cmidrule(lr){11-12}\cmidrule(lr){13-14}

& & & &{cSDR} &{uSDR} &{cSDR} &{uSDR} &{cSDR} &{uSDR} &{cSDR} &{uSDR} &{cSDR} &{uSDR}  \\

\midrule

&\textcolor{gray}{SIMO stereo BSRNN$^{\dagger}$~\cite{simo_stereo_bsrnn}} &\textcolor{gray}{Band-split (57)~\cite{simo_stereo_bsrnn}} &107.7 M &\textcolor{gray}{9.73} &\textcolor{gray}{10.27} &\textcolor{gray}{7.80} &\textcolor{gray}{7.61} &\textcolor{gray}{10.06} &\textcolor{gray}{9.83} &\textcolor{gray}{6.56} &\textcolor{gray}{6.50} &\textcolor{gray}{8.54} &\textcolor{gray}{8.55} \\

&\textcolor{gray}{TS-BSMAMBA2$^{\ddagger}$~\cite{mamba2_mss}} &\textcolor{gray}{Band-split (57)~\cite{simo_stereo_bsrnn}} &35.5 M &\textcolor{gray}{10.57} &\textcolor{gray}{10.60} &\textcolor{gray}{8.88} &\textcolor{gray}{7.47} &\textcolor{gray}{10.34} &\textcolor{gray}{10.07} &\textcolor{gray}{8.45} &\textcolor{gray}{6.69} &\textcolor{gray}{9.56} &\textcolor{gray}{8.71} \\

&\textcolor{gray}{BS-Roformer$^{\dagger}$~\cite{bsroformer}} &\textcolor{gray}{Band-split (62)~\cite{bsroformer}} &72.2 M &\textcolor{gray}{10.66} &\textcolor{gray}{\text{-}} &\textcolor{gray}{11.31} &\textcolor{gray}{\text{-}} &\textcolor{gray}{9.49} &\textcolor{gray}{\text{-}} &\textcolor{gray}{7.73} &\textcolor{gray}{\text{-}} &\textcolor{gray}{9.80} &\textcolor{gray}{\text{-}} \\

&BS-Locoformer (Medium)~\cite{saijo2025mixit} &Band-split (62)~\cite{bsroformer} &52.7 M &9.66 &9.83 &9.16 &8.14 &10.24 &10.45 &7.07 &6.57 &9.04 &8.75  \\

\midrule

\texttt{A1} &BS-Locoformer (Small) &Band-split (64)~\cite{bandit} &34.7 M &8.88 &9.01 &9.31 &8.21 &10.27 &9.93 &6.43 &5.87 &8.72 &8.26  \\

\texttt{A2} &SFC-Locoformer (Small) &SFC-Mamba (64) &5.1 M &8.97 &9.29 &9.25 &8.29 &10.36 &10.51 &6.97 &6.45 &8.86 &8.63  \\

\texttt{A3} &SFC-Locoformer (Small) &SFC-CA (64) &5.8 M &9.39 &9.61 &9.80 &8.70 &10.72 &10.75 &7.17 &6.71 &9.27 &8.95  \\

\midrule

\texttt{B1} &BS-Locoformer (Medium) &Band-split (64)~\cite{bandit} &55.5 M  &9.44 &9.59 &10.50 &8.90 &10.76 &10.43 &7.00 &6.24 &9.42 &8.79  \\

\texttt{B2} &SFC-Locoformer (Medium) &SFC-Mamba (64) &15.2 M &9.52 &9.63 &10.19 &8.93 &11.13 &10.05 &7.40 &6.90 &9.56 &9.13  \\

\texttt{B3} &SFC-Locoformer (Medium) &SFC-CA (64) &16.0 M &\bfseries 9.94 &\bfseries 10.18 &\bfseries 10.85 &\bfseries 9.15 &\bfseries 11.25 &\bfseries 11.08 &\bfseries 7.77 &\bfseries 7.12 &\bfseries 9.95 &\bfseries 9.38  \\ \hdashline

\textcolor{gray}{\texttt{B4}} &\textcolor{gray}{BS-Locoformer$^{\dagger}$ (Medium)} &\textcolor{gray}{Band-split (64)~\cite{bandit}} &\textcolor{gray}{55.5 M} &\textcolor{gray}{9.77} &\textcolor{gray}{10.02} &\textcolor{gray}{10.67} &\textcolor{gray}{9.28} &\textcolor{gray}{11.02} &\textcolor{gray}{10.94} &\textcolor{gray}{7.56} &\textcolor{gray}{6.81} &\textcolor{gray}{9.75} &\textcolor{gray}{9.26} \\

\bottomrule

\end{tabular}
}

\end{table*}

\begin{table*}[t]
\centering
\sisetup{
detect-weight, %
mode=text, %
tight-spacing=true,
round-mode=places,
round-precision=1,
table-format=2.1,
table-number-alignment=center
}
\caption{
    SNR and SISNR [dB] on DnR test set.
    Numbers in `Encoder/Decoder` cells are number of bands, where BandIt model, \texttt{C*}, and \texttt{D*} use \texttt{Musical} split.
    $\dagger$: Dynamic mixing is not used in training.
    $\ddagger$: Peformance reported in~\cite{bandit}.
}

\label{table:results_dnr}
\resizebox{\linewidth}{!}{
\begin{tabular}{lllc*{10}{S}}

\toprule

\multirow{2}{*}[-1.3ex]{\texttt{ID}} &\multirow{2}{*}[-1.3ex]{\shortstack{Model}} &\multirow{2}{*}[-1.3ex]{\shortstack{Encoder/Decoder}} &\multirow{2}{*}[-1.3ex]{\shortstack{Params.}} &\multicolumn{2}{c}{Speech} &\multicolumn{2}{c}{Music} &\multicolumn{2}{c}{Sfx} &\multicolumn{2}{c}{Average} \\
\cmidrule(lr){5-6}\cmidrule(lr){7-8}\cmidrule(lr){9-10}\cmidrule(lr){11-12}

& & & &{SNR} &{SISDR} &{SNR} &{SISDR} &{SNR} &{SISDR} &{SNR} &{SISDR}  \\

\midrule

&MRX-C$^\dagger$~\cite{dnr} &- &- &\text{-} &12.6 &\text{-} &4.6 &\text{-} &6.1 &\text{-} &7.8  \\

&Hybrid Demucs (v3)$^{\dagger \ddagger}$~\cite{hybrid_demucs} &- &83.6 M &13.6 &13.4 &6.0 &4.7 &7.2 &6.1 &8.9 &8.1  \\

&BandIt$^\dagger$~\cite{bandit} &Band-split (64)~\cite{bandit} &37.0 M &15.7 &15.6 &8.7 &8.0 &9.8 &8.2 &11.4 &10.9  \\

\midrule

\texttt{C1} &BS-Locoformer (Small) &Band-split (64)~\cite{bandit} &34.7 M &15.62 &15.52 &8.79 &8.27 &9.79 &9.09 &11.40 &10.96  \\

\texttt{C2} &SFC-Locoformer (Small) &SFC-Mamba (64) &5.1 M &15.68 &15.62 &9.00 &8.46 &9.99 &9.29 &11.56 &11.12  \\

\texttt{C3} &SFC-Locoformer (Small) &SFC-CA (64) &5.8 M &15.93 &15.86 &9.34 &8.83 &10.17 &9.449 &11.78 &11.38  \\

\midrule

\texttt{D1} &BS-Locoformer (Medium) &Band-split (64)~\cite{bandit} &55.5 M &16.15 &16.09 &9.36 &8.93 &10.29 &9.67 &11.93 &11.57  \\

\texttt{D2} &SFC-Locoformer (Medium) &SFC-Mamba (64) &15.2 M &16.24 &16.18 &9.55 &9.12 &10.45 &9.77 &12.08 &11.69  \\

\texttt{D3} &SFC-Locoformer (Medium) &SFC-CA (64) &16.0 M &\bfseries 16.38 &\bfseries 16.32 &\bfseries 9.70 &\bfseries 9.28 &\bfseries 10.56 &\bfseries 9.88 &\bfseries 12.21 &\bfseries 11.83  \\ \hdashline

\texttt{D4} &SFC-Locoformer (Medium)$^\dagger$ &SFC-CA (64) &16.0 M &16.2 &16.1 &9.2 &8.7 &10.1 &9.4 &11.8 &11.4  \\

\bottomrule

\end{tabular}
}

\end{table*}

In the following, we compare the performance of SFC and the BS module on the MUSDB18-HQ and DnR datasets.
Using MUSDB18-HQ, we also report ablation study results on the importance of inductive bias, the effect of different query types, and the performance when varying the number of bands $K$.
In addition, we present an analysis of the factors that contribute to the performance improvement of SFC.

\subsection{Main results on MSS}
\label{ssec:main results_mss}

\begin{figure}[t]
  \centering
  \includegraphics[width=0.75\linewidth]{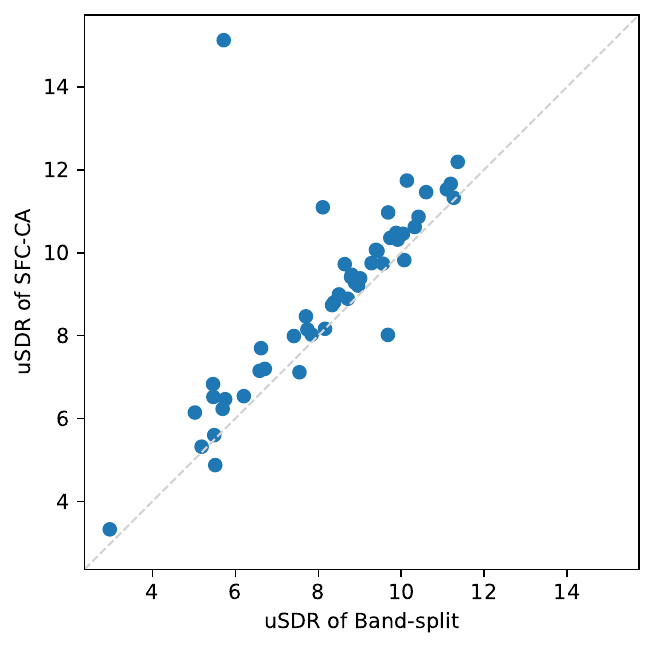}
  \caption{
    Scatter plot of uSDR [dB] values.
    Horizontal and vertical axes denote performances of small model with BS (\texttt{A1}) and SFC-CA (\texttt{A3}), respectively.
    }
    \label{fig:usdr_scatter_plot}
\end{figure}

Table~\ref{table:results_musdb} presents the cSDR and uSDR on the MUSDB18-HQ test set for both the small and medium models.
Scores reported in previous studies are also included; however, results from models that either did not explicitly specify their validation split or employed a custom split are shown in gray.

From Table~\ref{table:results_musdb}, we observe that the proposed SFC, whether implemented with Mamba or with cross-attention, outperforms the BS module in terms of average performance, regardless of separator size (small or medium).
In particular, SFC-CA consistently surpasses the other two across all instruments.
Notably, the medium model with SFC-CA outperforms existing state-of-the-art systems (\texttt{B3}).
To examine whether the performance improvement achieved by SFC-CA is statistically significant, we conducted paired t-tests comparing BS and SFC-CA (\texttt{A1} vs. \texttt{A3} and \texttt{B1} vs. \texttt{B3}).
For both the small and medium models, the p-values were approximately 0.001, and the effect size (Cohen’s d) was around 0.48, indicating a noticeable improvement in performance.
In addition, a comparison of the sample-wise uSDRs on the test set between BS (\texttt{A1}) and SFC-CA (\texttt{A3}), shown in Fig.~\ref{fig:usdr_scatter_plot}, demonstrates that SFC-CA outperformed BS on most of the tracks in the test set.
Comparison using medium models showed a similar trend.

The fact that SFC-CA substantially outperforms SFC-Mamba suggests that attention mechanisms, which can process the entire sequence simultaneously, are more effective than recurrent modeling in our case (\texttt{A2} vs. \texttt{A3} and \texttt{B2} vs. \texttt{B3}).
Nevertheless, Mamba still outperforms the BS module, indicating the potential of the recurrent variant.
Leveraging more advanced recurrent or linear attention techniques~\cite{mamba2, gated_delta_networks} may improve performance, but we leave it as future work.

Furthermore, comparing \texttt{A3} and \texttt{B1}, we find that the small model with the SFC-CA achieves performance comparable to the medium model with the BS module (lower cSDR but higher uSDR).
This result suggests that the design of the encoder and decoder has a significant impact on the final separation performance. While most prior research has primarily focused on improving the separator, our findings indicate that the design of the encoder and decoder may represent an important direction for future research.

\subsection{Main results on CASS}
\label{ssec:main results_cass}

Table~\ref{table:results_dnr} reports the SNR and SISDR on the DnR test set.
Several scores reported in prior studies are also included.
While we trained most models with dynamic mixing (\texttt{C1}-\texttt{C3} and \texttt{D1}-\texttt{D3}), we also trained a model without dynamic mixing (\texttt{D4}), to deliver a fairer comparison with prior studies.

In the CASS task, we observed trends consistent with those in the MSS task.
The proposed SFC consistently outperformed the BS module regardless of model size, with the cross-attention variant, SFC-CA, achieving the best performance.
Although the absolute improvement was somewhat smaller than in MSS, this result suggests that the proposed methods are particularly effective when separating sources with highly distinct frequency characteristics, such as musical instruments.
Nevertheless, in CASS as well, SFC consistently outperformed the BS module across all stems, confirming their effectiveness.
Our best model, SFC-Locoformer Medium, trained without dynamic mixing, outperformed existing state-of-the-art models, demonstrating the effectiveness of the proposed model.

\begin{figure}[t]
  \centering

  \includegraphics[width=\linewidth]{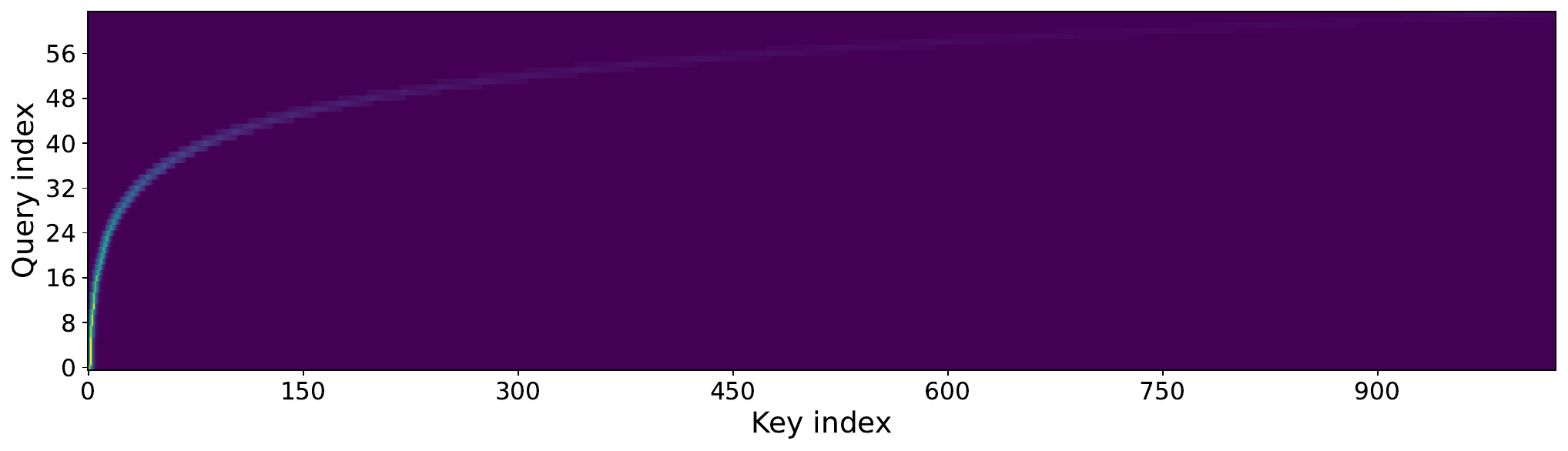}
  \includegraphics[width=\linewidth]{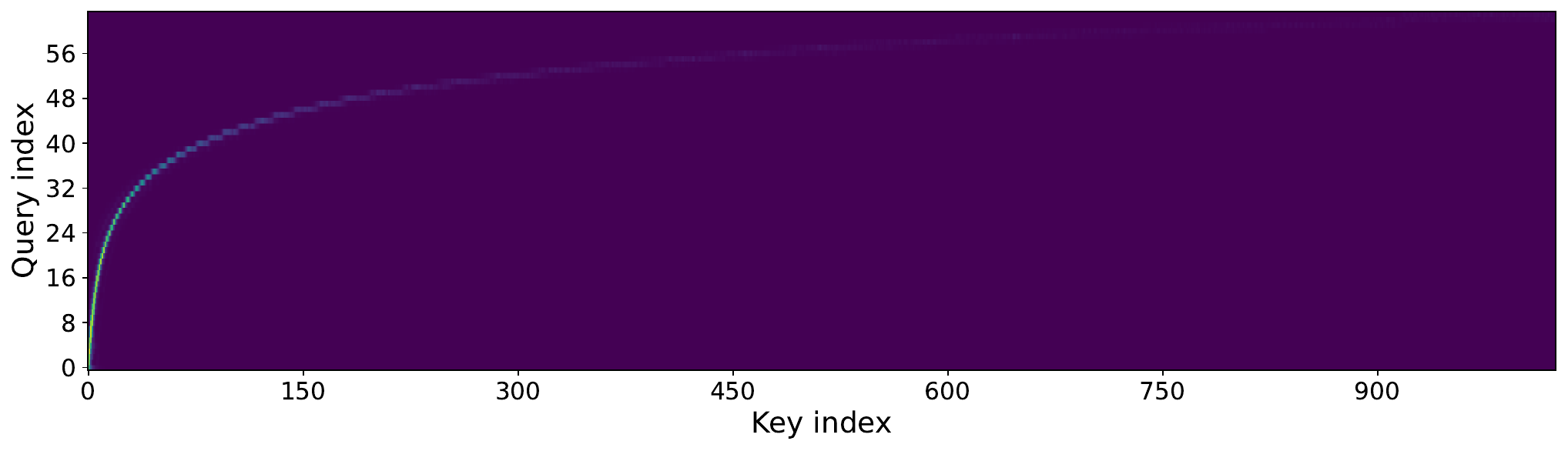}
  \includegraphics[width=\linewidth]{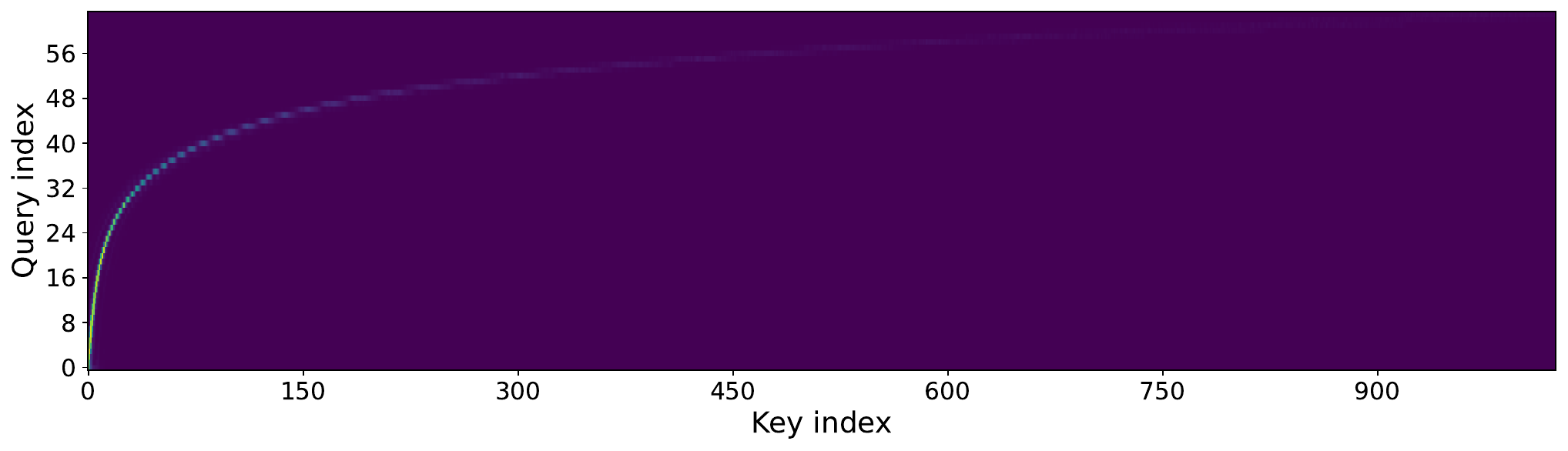}
  \includegraphics[width=\linewidth]{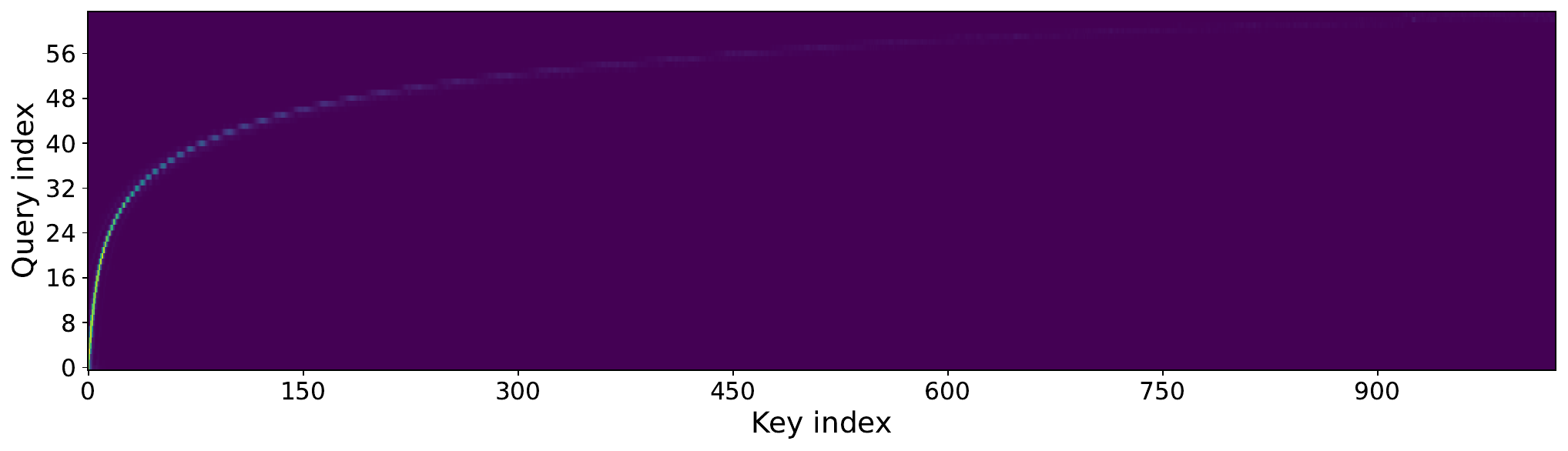}
  \includegraphics[width=\linewidth]{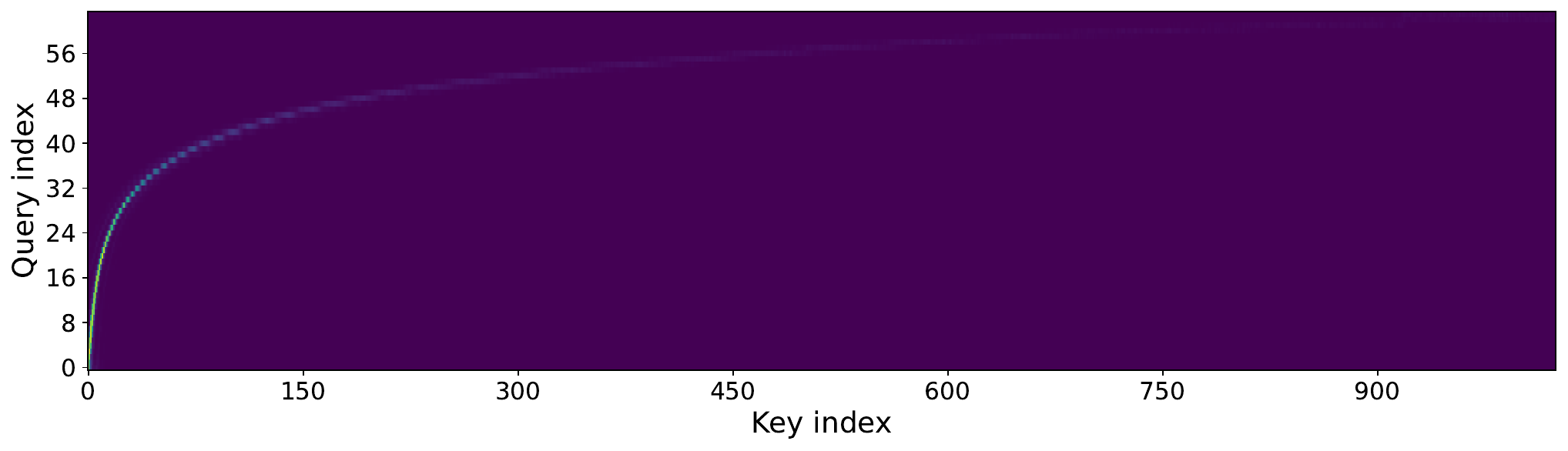}
  \caption{
    SFC-CA encoder's default positional bias $\bm{P}^{\E}$ defined in Eq.~(\ref{eq:position_bias}) (top) and learned positional bias of \texttt{E10} model in Table~\ref{table:ablation_inductive_bias} in each head (bottom four).
    For better visualization, we show $\mathrm{Softmax}(\bm{P}^{\E})$ instead of raw $\bm{P}^{\E}$. %
    }
    \label{fig:pos_bias}
\end{figure}

\begin{table*}[t]
\centering
\sisetup{
detect-weight, %
mode=text, %
tight-spacing=true,
round-mode=places,
round-precision=2,
table-format=2.2,
table-number-alignment=center
}
\caption{
    Ablation study on importance of inductive bias in each module. Small model with $K=64$ is evaluated on MUSDB18-HQ test set.
    In each type of encoder/decoder, methods above the dashed line do not include explicit inductive bias, while those below the dashed line do.
    $\bm{P}_{\mathrm{init}}$ in SFC-CA stands for the initial value of position bias.
}

\label{table:ablation_inductive_bias}
\resizebox{\linewidth}{!}{
\begin{tabular}{lll*{12}{S}}

\toprule

\multirow{2}{*}[-1.3ex]{\texttt{ID}} &\multirow{2}{*}[-1.3ex]{\shortstack{Enc./Dec.}} &\multirow{2}{*}[-1.3ex]{\shortstack{Config}} &\multicolumn{2}{c}{Vocals} &\multicolumn{2}{c}{Bass} &\multicolumn{2}{c}{Drums} &\multicolumn{2}{c}{Other} &\multicolumn{2}{c}{Average} \\
\cmidrule(lr){4-5}\cmidrule(lr){6-7}\cmidrule(lr){8-9}\cmidrule(lr){10-11}\cmidrule(lr){12-13}

& & &{cSDR} &{uSDR} &{cSDR} &{uSDR} &{cSDR} &{uSDR} &{cSDR} &{uSDR} &{cSDR} &{uSDR}  \\

\midrule

\texttt{E1} &Band-split &BS config: \texttt{Full} &8.15 &8.26 &8.60 &7.64 &9.02 &9.46 &6.25 &5.80 &8.01 &7.79 \\ \hdashline

\texttt{E2} &Band-split &BS config: \texttt{Musical}~\cite{bandit} &\bfseries 8.88 &\bfseries 9.01 &\bfseries 9.31 &\bfseries 8.21 &\bfseries 10.27 &\bfseries 9.93 &\bfseries 6.43 &\bfseries 5.87 &\bfseries 8.72 &\bfseries 8.26  \\

\midrule

\texttt{E3} &SFC-Mamba &Query-pos: Tail &3.44 &4.41 &3.79 &3.46 &4.49 &5.10 &2.82 &3.00 &3.64 &3.99 \\ \hdashline

\texttt{E4} &SFC-Mamba &Query-pos: Band start-end &8.84 &\bfseries 9.34 &9.06 &8.10 &\bfseries 10.44 &10.44 &6.72 &6.37 &8.77 &8.57 \\

\texttt{E5} &SFC-Mamba &Query-pos: Band middle &\bfseries 8.97 &9.29 &\bfseries 9.25 &\bfseries 8.29 &10.36 &\bfseries 10.51 &\bfseries 6.97 &\bfseries 6.45 &\bfseries 8.86 &\bfseries 8.63  \\

\midrule

\texttt{E6} &SFC-CA &Fixed $\gamma$, Fixed $\bm{P}$, $\bm{P}_{\mathrm{init}}=0$ &5.11 &5.52 &5.01 &4.86 &6.38 &7.00 &3.41 &3.61 &4.98 &5.25 \\

\texttt{E7} &SFC-CA &Fixed $\gamma$, Learn $\bm{P}$,~$\bm{P}_{\mathrm{init}}=0$ &8.48 &8.78 &7.50 &6.81 &9.14 &9.48 &5.98 &5.67 &7.77 &7.68  \\ \hdashline

\texttt{E8} &SFC-CA &Fixed $\gamma$, Fixed $\bm{P}$,~$\bm{P}_{\mathrm{init}}$: Eq.(\ref{eq:position_bias}) &9.21 &9.59 &9.45 &8.36 &10.26 &10.55 &7.39 &6.43 &9.08 &8.73  \\

\texttt{E9} &SFC-CA &Learn $\gamma$, Fixed $\bm{P}$,~$\bm{P}_{\mathrm{init}}$: Eq.(\ref{eq:position_bias}) &9.31 &9.50 &\bfseries 10.12 &8.68 &10.59 &10.50 &6.92 &6.31 &9.23 &8.75 \\

\texttt{E10} &SFC-CA &Fixed $\gamma$, Learn $\bm{P}$,~$\bm{P}_{\mathrm{init}}$: Eq.(\ref{eq:position_bias}) &9.39 &9.61 &9.80 &\bfseries 8.70 &\bfseries 10.72 &\bfseries 10.75 &\bfseries 7.17 &\bfseries 6.71 &\bfseries 9.27 &\bfseries 8.95 \\

\texttt{E11} &SFC-CA &Learn $\gamma$, Learn $\bm{P}$,~$\bm{P}_{\mathrm{init}}$: Eq.(\ref{eq:position_bias}) &\bfseries 9.43 &\bfseries 9.70 &9.50 &8.46 &10.32 &10.52 &7.12 &6.64 &9.09 &8.83 \\

\bottomrule

\end{tabular}
}

\end{table*}

\begin{table*}[t]
\centering
\sisetup{
detect-weight, %
mode=text, %
tight-spacing=true,
round-mode=places,
round-precision=2,
table-format=2.2,
table-number-alignment=center
}
\caption{
    Ablation study on query choice.
    Small model with $K=64$ are evaluated on MUSDB18-HQ test set.
}

\label{table:ablation_query}
\resizebox{\linewidth}{!}{
\begin{tabular}{lll*{12}{S}}

\toprule

\multirow{2}{*}[-1.3ex]{\texttt{ID}} &\multirow{2}{*}[-1.3ex]{\shortstack{Encoder/Decoder}} &\multirow{2}{*}[-1.3ex]{\shortstack{Query}} &\multicolumn{2}{c}{Vocals} &\multicolumn{2}{c}{Bass} &\multicolumn{2}{c}{Drums} &\multicolumn{2}{c}{Other} &\multicolumn{2}{c}{Average} \\
\cmidrule(lr){4-5}\cmidrule(lr){6-7}\cmidrule(lr){8-9}\cmidrule(lr){10-11}\cmidrule(lr){12-13}

& & &{cSDR} &{uSDR} &{cSDR} &{uSDR} &{cSDR} &{uSDR} &{cSDR} &{uSDR} &{cSDR} &{uSDR}  \\

\midrule

\texttt{F1} &SFC-Mamba &\texttt{Adaptive} &\bfseries 8.97 &\bfseries 9.29 &\bfseries 9.25 &\bfseries 8.29 &\bfseries 10.36 &10.51 &\bfseries 6.97 &\bfseries 6.45 &\bfseries 8.86 &\bfseries 8.63 \\

\texttt{F2} &SFC-Mamba &\texttt{Learnable} &8.73 &9.07 &8.90 &7.94 &10.10 &\bfseries 10.59 &6.14 &6.02 &8.47 &8.41 \\ \hdashline

\texttt{F3} &SFC-CA &\texttt{Adaptive} &\bfseries 9.46 &\bfseries 9.75 &9.26 &8.40 &\bfseries 10.86 &10.70 &\bfseries 7.17 &6.59 &9.19 &8.86 \\

\texttt{F4} &SFC-CA &\texttt{Learnable} &9.39 &9.61 &\bfseries 9.80 &\bfseries 8.70 &10.72 &\bfseries 10.75 &\bfseries 7.17 &\bfseries 6.71 &\bfseries 9.27 &\bfseries 8.95 \\

\bottomrule

\end{tabular}
}

\end{table*}

\subsection{Is inductive bias really important?}
\label{ssec:ablation_inductive_bias}

As discussed in Section~\ref{sec:proposed_methods}, the proposed SFC module was designed to incorporate a psychoacoustically motivated inductive bias, inspired by the BS module.
To examine the importance of this inductive bias, we conducted an ablation study and report the results in Table~\ref{table:ablation_inductive_bias}.

First, to examine whether the inductive bias in the BS module is indeed effective, we trained a model in which all sub-encoders were allowed to access all frequency bins (i.e., $\bandstart = 1, \bandend = F$ for all $k$).
We denote this configuration as \texttt{E1}, (termed as \texttt{Full} split) in Table~\ref{table:ablation_inductive_bias}.
In principle, such a model should be able to learn behavior similar to that of a model with an inductive bias.
However, comparison between \texttt{E1} and \texttt{E2} shows that it fails to do so, confirming that the inductive bias introduced in the BS module is indeed effective.

Second, we examined the effect of interleaving strategies in Mamba.
In addition to the \texttt{Band start-end} and \texttt{Band middle} strategies described in Section~\ref{sssec:interleaving_strategy}, we also trained a model in which the query $\bm{Q}$ was simply concatenated to the tail of the feature sequence $\bm{Z}$, which does not introduce any inductive bias (denoted as \texttt{E3}).
As shown in the table, this model performs significantly worse than the other two interleaving strategies, demonstrating the effectiveness of introducing inductive bias through query insertion positions.
Among the three, \texttt{Band middle} achieved the best performance.

Finally, we investigated the effect of the proposed positional bias in cross-attention.
We compared models trained with and without the positional bias, and further examined variants in which either the slope $\gamma$ or the positional bias matrix $\bm{P}$ was learnable.
The results indicate that models with positional bias (\texttt{E8}-\texttt{E11}) significantly outperform those without it (\texttt{E6}), demonstrating the effectiveness of introducing inductive bias through positional bias.
Moreover, learning $\bm{P}$ provides better performance than learning $\gamma$, which is consistent with the discussion in Section~\ref{sssec:crossattn_with_inductive_bias}.
However, no further improvement was observed when both were learned simultaneously (\texttt{E11}).
This is likely because learning $\bm{P}$ and learning $\gamma$ have similar effects: both make the positional bias peakier than the default setting defined in Eq.~(\ref{eq:position_bias}).
Figure~\ref{fig:pos_bias} illustrates the encoder’s default positional bias given by Eq.~(\ref{eq:position_bias}) (top) and the learned positional biases of the \texttt{E10} model after applying the softmax operation (bottom four). The figure shows that the learned positional biases are noticeably peakier than the default one.
We also examined the learned slope parameter $\gamma$ in the \texttt{E9} model, which is set to 1 by default. The learned values for the four heads were 3.89, 3.51, 3.83, and 3.34, respectively. This result likewise indicates that $\gamma \bm{P}^{\E}$ becomes peakier than in the default configuration.
Nevertheless, whereas learning $\gamma$ only adjusts the slope of the positional bias, learning $\bm{P}$ enables the model to directly control where to attend (see Section~\ref{sssec:crossattn_with_inductive_bias} for further discussion). We believe that this increased flexibility explains why \texttt{E10} outperforms \texttt{E9}.
Comparing \texttt{E7} and \texttt{E10}, both of which employ learnable positional bias but with different initializations, \texttt{E10} outperforms \texttt{E7}.
This result suggests that models without explicit inductive bias struggle to match the performance of those with such bias, consistent with the observations in the BS module.
These three ablation studies consistently demonstrate that the inductive bias is crucial for compressing frequency information, supporting the validity of the proposed method.

\subsection{Query choice}
\label{ssec:ablation_query}

As described in Sections~\ref{sssec:learnable_query} and \ref{sssec:input_query}, we consider two types of queries, \texttt{Learnable} and \texttt{Adaptive}.
Table~\ref{table:ablation_query} shows the uSDR and cSDR scores on the MUSDB18-HQ test set when changing the query configuration.

The results demonstrate that Mamba works better with the \texttt{Adaptive} query, likely because it directly scans the query to compress or restore the sequence, as discussed in the last paragraph of Section~\ref{sssec:input_query}.
In contrast, cross-attention performs slightly better with the \texttt{Learnable} query.
While it works well even with \texttt{Adaptive} query, it needs a slight but non-negligible additional computational cost.
We thus conclude that \texttt{Learnable} query is the better choice for cross-attention.

\begin{table*}[t]
\centering
\sisetup{
detect-weight, %
mode=text, %
tight-spacing=true,
round-mode=places,
round-precision=2,
table-format=2.2,
table-number-alignment=center
}
\caption{
    Performance of small model when changing number of bands $K$ on MUSDB18-HQ test set.
}

\label{table:ablation_num_bands}
\resizebox{\linewidth}{!}{
\begin{tabular}{llc*{13}{S}}

\toprule

\multirow{2}{*}[-1.3ex]{\texttt{ID}} &\multirow{2}{*}[-1.3ex]{\shortstack{Encoder/Decoder}} &\multirow{2}{*}[-1.3ex]{\shortstack{\#Bands \\$K$}} &\multicolumn{2}{c}{Vocals} &\multicolumn{2}{c}{Bass} &\multicolumn{2}{c}{Drums} &\multicolumn{2}{c}{Other} &\multicolumn{3}{c}{Average} \\
\cmidrule(lr){4-5}\cmidrule(lr){6-7}\cmidrule(lr){8-9}\cmidrule(lr){10-11}\cmidrule(lr){12-14}

& & &{cSDR} &{uSDR} &{cSDR} &{uSDR} &{cSDR} &{uSDR} &{cSDR} &{uSDR} &{cSDR} &{uSDR} &{SpecSNR}  \\

\midrule

\texttt{G1} &Band-split &32 &7.78 &7.87 &7.80 &7.33	&8.69 &8.98	&5.75 &4.96	&7.51 &7.28 &9.09 \\

\texttt{G2} &SFC-Mamba &32 &8.13 &8.46 &8.53 &7.87 &9.62 &9.88 &6.00 &5.61 &8.07 &7.95 &9.95 \\

\texttt{G3} &SFC-CA &32 &\bfseries 8.58 &\bfseries 8.85 &\bfseries 9.08 &\bfseries 8.35 &\bfseries 10.37 &\bfseries 10.25 &\bfseries 6.32 &\bfseries 5.85 &\bfseries 8.56 &\bfseries 8.32 &\bfseries 10.30 \\

\midrule

\texttt{G4} &Band-split &48 &8.49 &8.47	&8.59 &8.05	&9.93 &9.82	&6.04 &5.56	&8.26 &7.98 &9.85 \\

\texttt{G5} &SFC-Mamba &48 &8.46 &8.94 &9.13 &8.10 &10.10 &10.22 &6.01 &5.93 &8.43 &8.30 &10.32 \\

\texttt{G6} &SFC-CA &48 &\bfseries 8.96 &\bfseries 9.32 &\bfseries 9.75 &\bfseries 8.45 &\bfseries 10.68 &\bfseries 10.71 &\bfseries 6.85 &\bfseries 6.39 &\bfseries 9.06 &\bfseries 8.72 &\bfseries 10.75 \\

\midrule

\texttt{G7} &Band-split &64 &8.88 &9.01 &9.31 &8.21 &10.27 &9.93 &6.43 &5.87 &8.72 &8.26 &10.17 \\

\texttt{G8} &SFC-Mamba &64 &8.97 &9.29 &9.25 &8.29 &10.36 &10.51 &6.97 &6.45 &8.86 &8.63 &10.69 \\

\texttt{G9} &SFC-CA &64 &\bfseries 9.39 &\bfseries 9.61 &\bfseries 9.80 &\bfseries 8.70 &\bfseries 10.72 &\bfseries 10.75 &\bfseries 7.17 &\bfseries 6.71 &\bfseries 9.27 &\bfseries 8.95 &\bfseries 10.97 \\

\bottomrule

\end{tabular}
}

\end{table*}

\subsection{Performance on higher compression ratio}
\label{ssec:ablation_bands}

To evaluate the proposed method with a higher compression ratio (i.e., with a smaller number of bands $K$), we trained the models with $K$ set to 32, 48, or 64, and report the performances in Table~\ref{table:ablation_num_bands}.
The results show that the proposed methods consistently outperform the BS module across all values of $K$.
Moreover, the performance degradation of SFC under higher compression ratios is smaller than that of the BS module, indicating that the proposed methods are particularly effective in high-compression settings.
Notably, SFC-CA with $K=32$ performs comparably to the BS module with $K=64$, highlighting again the importance of the encoder and decoder design in source separation models.
This trend is consistent in terms of the spectrogram reconstruction quality (SpecSNR, Eq.~(\ref{eq:specsnr})), which also supports that the proposed SFC module compresses and decompresses the spectral information well.

\subsection{Ablation study using RNN-based separator}
\label{ssec:ablation_gru}

\begin{table*}[t]
\centering
\sisetup{
detect-weight, %
mode=text, %
tight-spacing=true,
round-mode=places,
round-precision=2,
table-format=2.2,
table-number-alignment=center
}
\caption{
    Performance of GRU-based model when changing encoder and decoder.
}

\label{table:ablation_gru}
\resizebox{0.9\linewidth}{!}{
\begin{tabular}{ll*{12}{S}}

\toprule

\multirow{2}{*}[-1.3ex]{\texttt{ID}} &\multirow{2}{*}[-1.3ex]{\shortstack{Encoder/Decoder}} &\multicolumn{2}{c}{Vocals} &\multicolumn{2}{c}{Bass} &\multicolumn{2}{c}{Drums} &\multicolumn{2}{c}{Other} &\multicolumn{2}{c}{Average} \\
\cmidrule(lr){3-4}\cmidrule(lr){5-6}\cmidrule(lr){7-8}\cmidrule(lr){9-10}\cmidrule(lr){11-12}

& &{cSDR} &{uSDR} &{cSDR} &{uSDR} &{cSDR} &{uSDR} &{cSDR} &{uSDR} &{cSDR} &{uSDR}  \\

\midrule

\texttt{H1} &Band-split (64) &9.00 &9.27 &9.18 &8.18 &8.98 &9.25 &6.58 &5.85 &8.44 &8.14 \\

\texttt{H2} &SFC-Mamba (64) &9.27 &9.38 &9.32 &7.82 &9.29 &9.64 &7.02 &6.11	&8.73 &8.24 \\

\texttt{H3} &SFC-CA (64) &\bfseries 9.51 &\bfseries 9.63 &\bfseries 9.53 &\bfseries 8.32 &\bfseries 9.66 &\bfseries 9.83 &\bfseries 7.21 &\bfseries 6.14 &\bfseries 8.98 &\bfseries 8.48 \\

\bottomrule

\end{tabular}
}

\end{table*}

Although all preceding experiments employed the TF-Locoformer backbone, we additionally conducted an ablation study using an RNN-based backbone to examine whether the SFC remains effective across different separator architectures. 
Specifically, we replaced the TF-Locoformer blocks with eight stacks of dual-path GRU blocks, as in~\cite{bandit}.
The encoder and decoder configurations were identical to those of the SFC-Locoformer medium model, while the separator configuration followed~\cite{bandit}.

The results indicate that the proposed SFC module is also effective with RNN-based separators, suggesting that its benefits are largely independent of the underlying separator backbone. 
A similar performance trend to the Locoformer setting is observed: both SFC variants outperform the BS module, with SFC-CA achieving the best results.

\subsection{What contributes to performance improvement?}
\label{ssec:analysis}

\begin{figure*}[t]
  \centering

  \includegraphics[width=.2\linewidth]{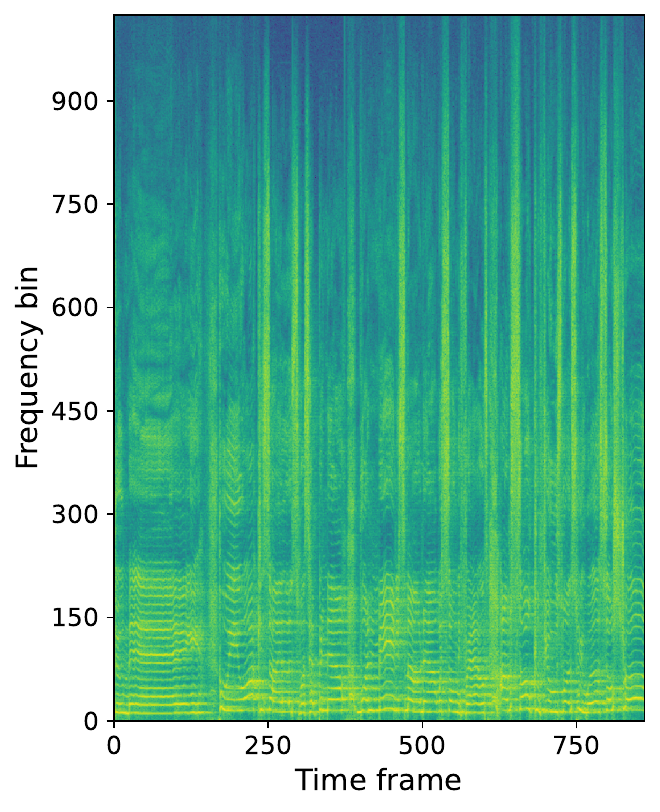}\hfill
  \includegraphics[width=.2\linewidth]{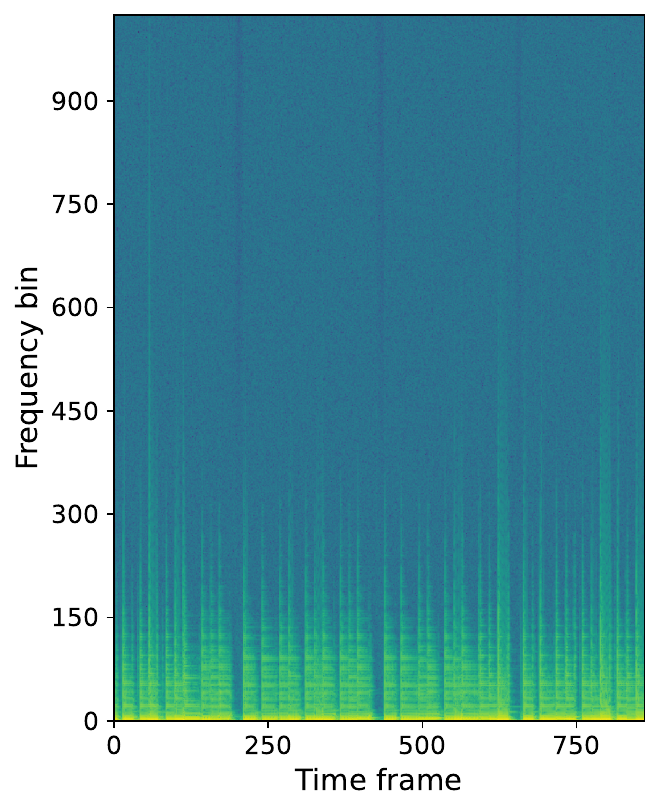}\hfill
  \includegraphics[width=.2\linewidth]{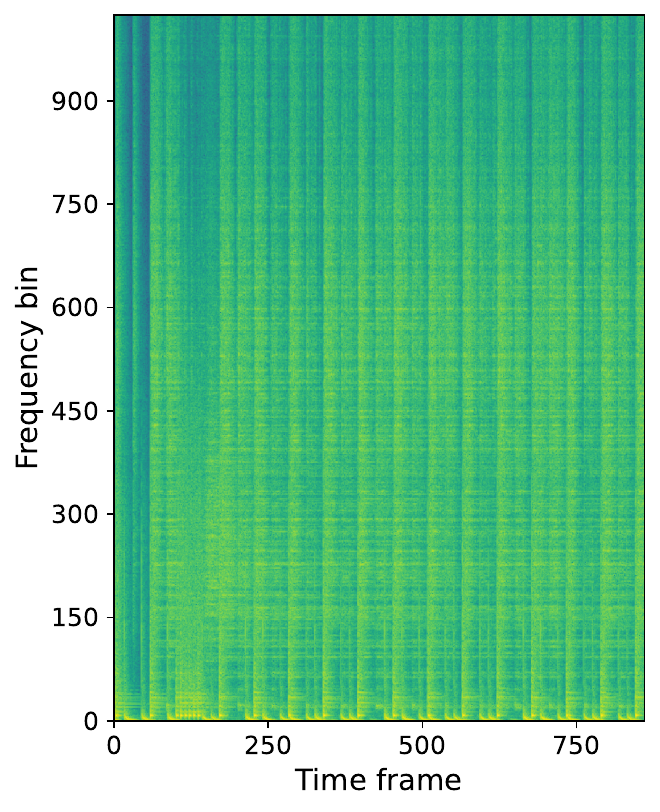}\hfill
  \includegraphics[width=.2\linewidth]{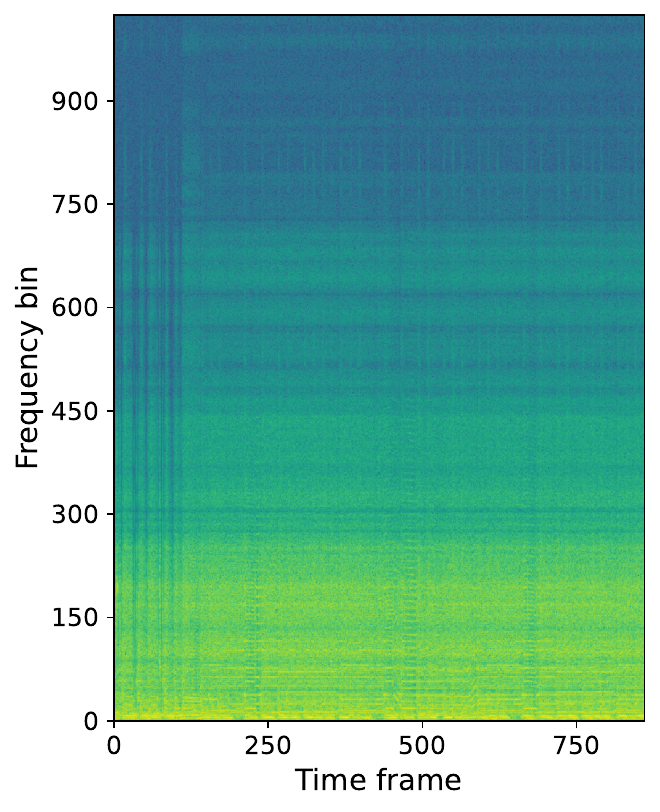}\hfill
  \includegraphics[width=.2\linewidth]{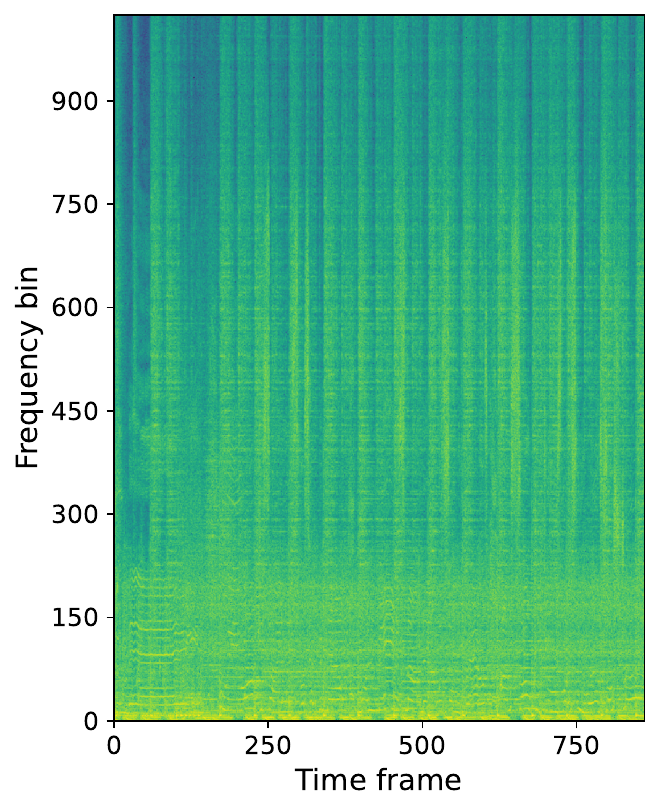}

  \includegraphics[width=.2\linewidth]{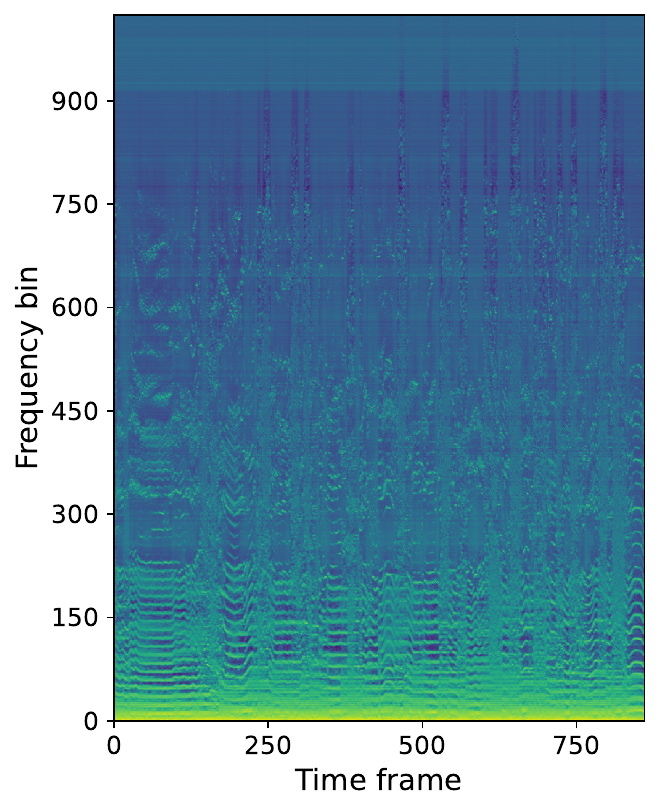}\hfill
  \includegraphics[width=.2\linewidth]{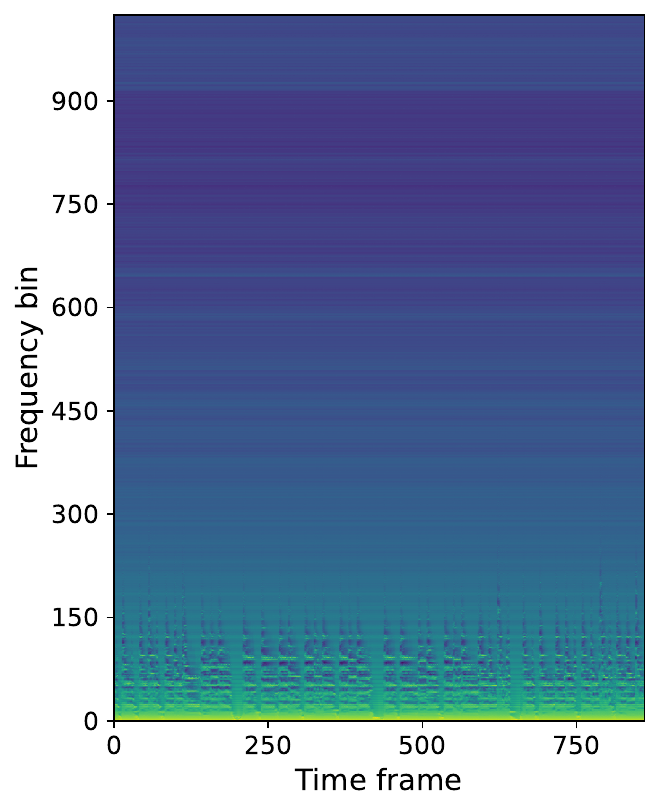}\hfill
  \includegraphics[width=.2\linewidth]{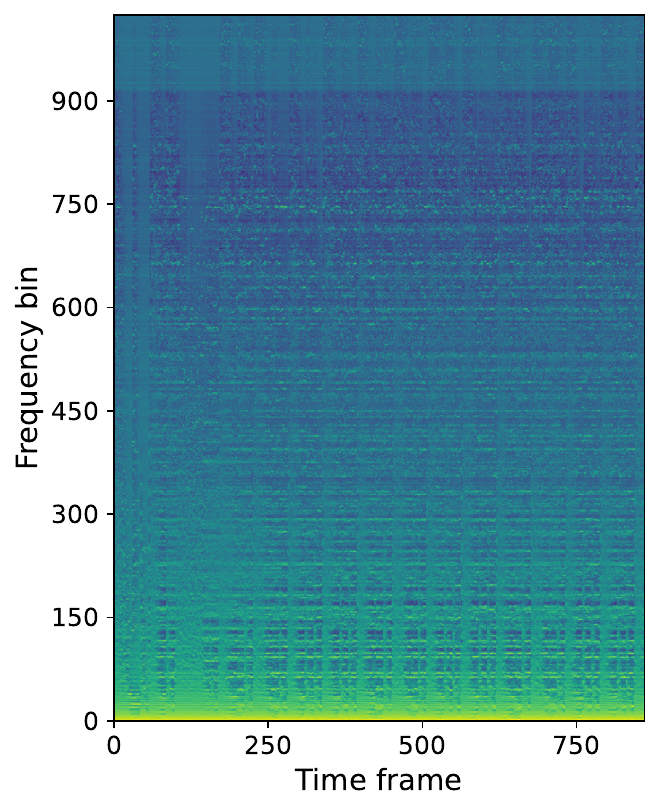}\hfill
  \includegraphics[width=.2\linewidth]{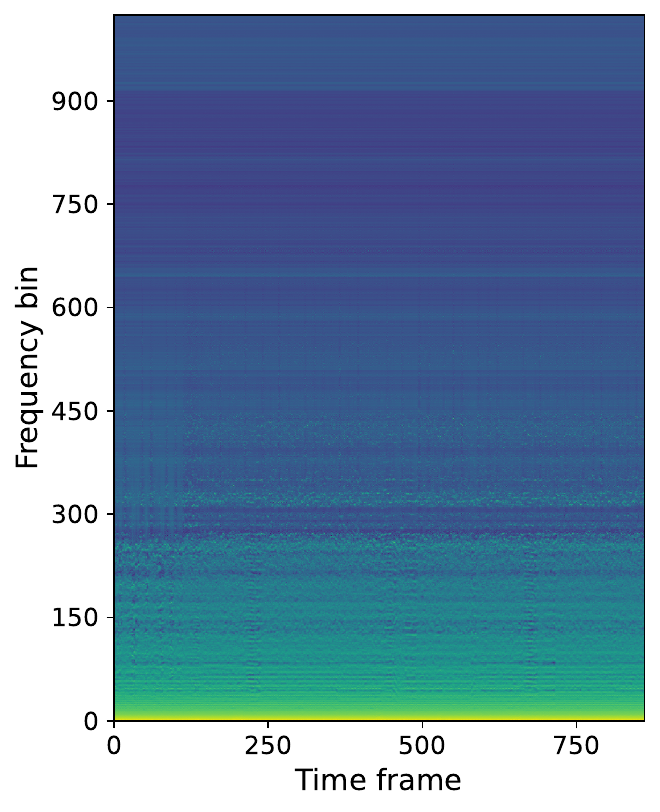}\hfill
  \includegraphics[width=.2\linewidth]{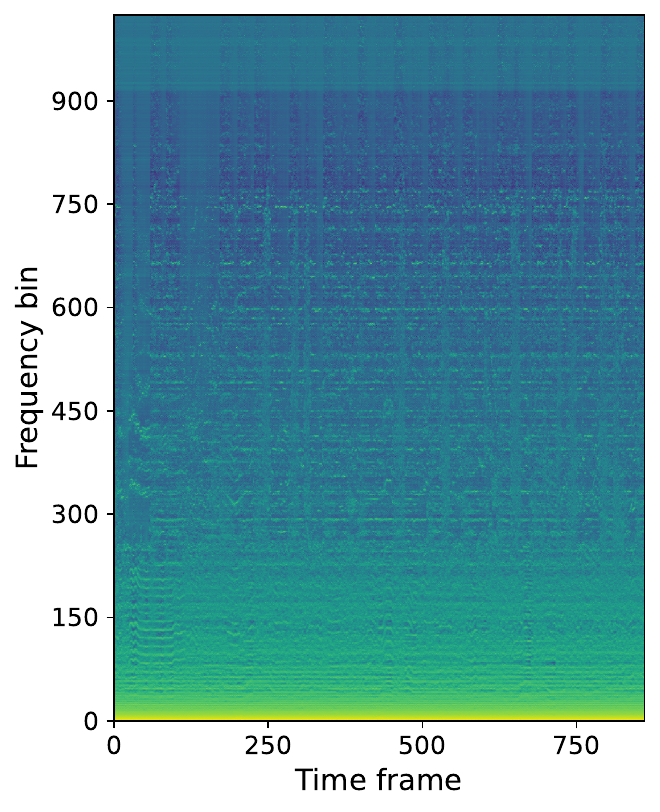}
  \caption{
    Log-magnitude spectrograms (top) and log-attention-weight spectrograms (bottom) of vocals, bass, drums, other, and mixture (from left to right). Note that the spectrograms correspond to the ground-truth signals, not the separated signals.
    }
    \label{fig:attention_spectrogram}
\end{figure*}

\begin{table}[t]
\centering
\sisetup{
detect-weight, %
mode=text, %
tight-spacing=true,
round-mode=places,
round-precision=2,
table-format=2.2,
table-number-alignment=center
}
\caption{
    Performance when not limiting (\texttt{I1}) and limiting (\texttt{I2}) receptive field of cross-attention.
    Small model with $K=64$ are evaluated on MUSDB18-HQ test set.
}

\label{table:ablation_receptive_field}
\resizebox{\linewidth}{!}{
\setlength{\tabcolsep}{3pt} %
\begin{tabular}{@{} lc*{12}{S} @{}} %

\toprule

\multirow{2}{*}[-1.3ex]{\texttt{ID}} &\multirow{2}{*}[-1.3ex]{\shortstack{Limit \\ rec. field}} &\multicolumn{2}{c}{Vocals} &\multicolumn{2}{c}{Bass} &\multicolumn{2}{c}{Drums} &\multicolumn{2}{c}{Other} &\multicolumn{2}{c}{Average} \\
\cmidrule(lr){3-4}\cmidrule(lr){5-6}\cmidrule(lr){7-8}\cmidrule(lr){9-10}\cmidrule(lr){11-12}

& &{cSDR} &{uSDR} &{cSDR} &{uSDR} &{cSDR} &{uSDR} &{cSDR} &{uSDR} &{cSDR} &{uSDR}  \\

\midrule

\texttt{I1} & &\bfseries 9.39 &9.61 &\bfseries 9.80 &\bfseries 8.70 &\bfseries 10.72 &\bfseries 10.75 &\bfseries 7.17 &\bfseries 6.71 &\bfseries 9.27 &\bfseries 8.95 \\

\texttt{I2} &\checkmark &9.34 &\bfseries 9.68 &9.58 &8.22 &10.48 &10.32 &7.11 &6.39 &9.13 &8.65 \\

\bottomrule

\end{tabular}
}

\end{table}

As discussed in Section~\ref{sec:proposed_methods}, the superior performance of the SFC module compared with the BS module is likely attributable to the fact that SFC is input-adaptive.
Another key difference, however, is that unlike the BS module, SFC does not explicitly restrict the receptive field\footnote{Note that, however, the receptive field is not unlimited. Under our experimental setting (state size 8), Mamba cannot retain all frequency information. Likewise, in cross-attention, the position bias becomes very small for distant frequency bins, which is essentially equivalent to restricting the receptive field to some extent~\cite{alibi_receptive_field}.}.
Thus, it is also possible that the difference in receptive field contributes to the performance gap.
To disentangle the impact of these two factors, we conducted two experiments: (i) performance evaluation of SFC when its receptive field was restricted in the same manner as the BS module, and (ii) verification of whether SFC truly operates in an input-adaptive manner.

\subsubsection{Performance with limited receptive field}
Since restricting the receptive field of Mamba is not straightforward, we used the SFC-CA for this experiment.
Specifically, for each band $k$, we applied an attention mask to mask out frequency bins not included in the corresponding band $G_k$.

Table~\ref{table:ablation_receptive_field} shows the results of SFC-CA without and with receptive field restriction.
The results show that a wider receptive field contributes to improving performance.
However, the model with the receptive field limitation (\texttt{I2}) still outperforms the BS module, implying that the improved performance of SFC cannot be explained solely by its wider receptive field.

\subsubsection{Input-adaptiveness of SFC}
Next, we examined whether SFC adaptively captures input-dependent information.
Since it is not easy to evaluate this aspect with Mamba, we analyzed the attention weights of the cross-attention in the encoder.
Because SFC is applied independently to each frame, we inspected whether input frequency patterns were captured on a frame-by-frame basis.
Applying cross-attention with $H$ heads to $T$ frames produces attention weights with the shape $T \times H \times K \times F$.
As we are interested in whether frequency patterns are captured at each frame, we averaged the attention weight over the head dimension $H$ and band dimension $K$, reducing the representation to $T \times F$, which we refer to as the \textit{attention-weight spectrogram}.

Fig.~\ref{fig:attention_spectrogram} shows the log-magnitude spectrograms of the input signals (top) and the log-attention-weight spectrograms obtained from the encoder (bottom).
To examine input adaptiveness, we present five cases: when inputting the ground-truth signals of vocals, bass, drums, and other (left four), which have distinct frequency characteristics, and when inputting their mixture (right).
Note that, for better visualization, the attention-weight spectrograms were converted to a logarithmic scale.
Interestingly, the attention-weight spectrograms exhibit structures similar to those of input spectrograms, which suggests that cross-attention effectively captures input-specific frequency structures.
These results indicate that SFC indeed operates in an input-adaptive manner, which we believe contributes to its strong performance.

\subsection{Training and inference costs}
\label{ssec:train_inf_costs}

\begin{table}[t]
\centering
\sisetup{
  detect-weight,
  mode=text,
  tight-spacing=true,
  table-number-alignment=center
}
\caption{
    Training and inference statistics with different encoders and decoders.
    TF-Locoformer blocks were used as separators.
    See Section~\ref{ssec:separator} and Section~\ref{ssec:enc_dec} for detailed setup of models.
    Training time indicates the duration of a training epoch (without validation steps) in seconds with a 6-second input.
}
\label{table:train_inf_costs}

\resizebox{\linewidth}{!}{%
\setlength{\tabcolsep}{3pt}
\begin{tabular}{@{} l l l c
  S[table-format=2.2]
  S[table-format=2.2]
  S[table-format=1.4] @{}} %
\toprule
\texttt{ID} & Size & Enc/Dec & Params.
& \multicolumn{1}{c}{Train.\ time (s)}
& \multicolumn{1}{c}{FLOPs (G/s)}
& \multicolumn{1}{c}{RTF} \\

\midrule

\texttt{J1} &  & BS        & 34.7M &21.8  &36.49  & 0.0017 \\
\texttt{J2} & Small & SFC-Mamba & 5.1M  &22.9  &40.14  & 0.0025 \\
\texttt{J3} &       & SFC-CA    & 5.8M  &18.3  &41.04  & 0.0018 \\

\midrule

\texttt{J4} &  & BS              &55.5M &32.3  &100.06  & 0.0031 \\
\texttt{J5} & Medium & SFC-Mamba &15.2M  &45.6  &108.82  & 0.0046 \\
\texttt{J6} &        & SFC-CA    &16.0M  &31.4  &110.37  & 0.0035 \\

\bottomrule

\end{tabular}%
}
\end{table}

Table~\ref{table:train_inf_costs} summarizes several statistics related to training and inference.
All experiments were conducted using PyTorch~2.6.
We used 8 NVIDIA H200 GPUs for training and 1 for inference\footnote{Note that training did not strictly require eight GPUs; it was feasible with one or two H200 GPUs. However, eight GPUs were used to accelerate the training process.}.
The training time corresponds to the duration of a single training epoch (110 training steps), measured in seconds, with a 6-second input.
It was computed by running 10 epochs (1100 training steps) and averaging the elapsed time.
FLOPs and real-time factor (RTF) were measured using 12-second inputs.
Here, FLOPs denote the number of floating-point operations per second, i.e., the total FLOPs divided by 12.
RTF was obtained by performing inference 500 times and computing the average value.

As shown in Table~\ref{table:train_inf_costs}, although SFC-CA introduces additional computational overhead compared with the BS module, it achieves faster training and comparable inference speed.
We note that FlashAttention was not used in SFC-CA because PyTorch~2.6 does not support positional bias in FlashAttention (at the time of writing).
This suggests that SFC-CA could achieve further speed improvements once such support becomes available.
Although SFC-Mamba requires fewer FLOPs than SFC-CA, it was significantly slower in both training and inference in our experiments.
These results indicate that SFC-CA is more favorable than SFC-Mamba under the considered settings.
Reducing the computational cost of SFC-CA without sacrificing performance remains an interesting direction for future work.
\section{Conclusion}
\label{sec:conclusion}

We proposed Spectral Feature Compression (SFC) to effectively compress and uncompress the spectral feature for the TF-domain dual-path source separation models.
We explore two variants of the proposed method, SFC-CA and SFC-Mamba.
In both cases, we introduce a simple yet effective way to incorporate an inductive bias, as in the BS module, making them suitable for frequency modeling.
Extensive experiments have shown that the proposed SFC module results in better separation performance than the BS module in the MSS and CASS tasks.
The SFC module consistently outperformed the BS module across multiple separator sizes and compression ratios.
Notably, the small separation model with SFC-CA performed on par with the medium one with the BS module.
We also analyzed the behavior of the SFC, and showed that it adaptively captures input-dependent frequency structure.
While the compression ratio of SFC was treated as a fixed parameter in this work, designing a mechanism that allows the compression ratio to be adjusted dynamically would be an interesting direction for future work.

\bibliographystyle{IEEEtran}
\bibliography{main}

\end{document}